\documentclass[12pt,preprint]{aastex}

\slugcomment{draft of \today}

\shorttitle{Nonthermal Radiation from Cluster Shocks}
\shortauthors{Kang \& Ryu}

\def\eg{{\it e.g.,~}}
\def\ie{{\it i.e.,~}}

\def\kms{~{\rm km~s^{-1}}}
\def\cm3{~{\rm cm^{-3}}}
\def\yr{~{\rm yr}}
\def\Myr{~{\rm Myr}}
\def\muG{~{\mu\rm G}}

\def\kpc{~{\rm kpc}}

\begin{document}
\title{CURVED RADIO SPECTRA OF WEAK CLUSTER SHOCKS}

\author{Hyesung Kang$^1$ and Dongsu Ryu$^2$}

\affil{$^1$Department of Earth Sciences, Pusan National University, Pusan 609-735, Republic of Korea: hskang@pusan.ac.kr\\
$^2$Department of Physics, UNIST, Ulsan 689-798, Korea: ryu@sirius.unist.ac.kr}

\begin{abstract}
{
In order to understand certain observed features of arc-like giant radio relics such as
the rareness, uniform surface brightness, and curved integrated spectra, 
we explore a diffusive shock acceleration (DSA) model for radio relics in which 
a spherical shock impinges on a magnetized cloud containing fossil relativistic electrons. 
}
Toward this end, we perform DSA simulations of spherical shocks 
with the parameters relevant for the Sausage radio relic in cluster CIZA J2242.8+5301,
and calculate the ensuing radio synchrotron emission from re-accelerated electrons.
{ 
Three types of fossil electron populations are considered: a delta-function like population with 
the shock injection momentum, a power-law distribution, and a power-law with an exponential cutoff.}
The surface brightness profile of radio-emitting postshock region and the volume-integrated 
radio spectrum are calculated and compared with observations.
We find that the observed width of the Sausage relic can be explained 
reasonably well by shocks with speed $u_s \sim 3\times 10^3 \kms$ and sonic Mach number $M_s \sim 3$.
These shocks produce curved radio spectra that steepen gradually over
$(0.1-10) \nu_{\rm br}$ with break frequency $ \nu_{\rm br}\sim 1$ GHz,
if the duration of electron acceleration is $\sim 60 - 80$ Myr. 
However, the abrupt increase of spectral index above $\sim 1.5$ GHz observed in
the Sausage relic seems to indicate that additional physical processes,
other than radiative losses, operate for electrons with $\gamma_e \gtrsim 10^4$.
\end{abstract}


\section{INTRODUCTION}

{
Radio relics are diffuse radio sources found in the outskirts of galaxy clusters and 
they are thought to trace synchrotron-emitting cosmic-ray (CR) electrons accelerated via
diffusive shock acceleration (DSA) at cluster shocks \citep[e.g.][]{ensslin98, bagchi06, vanweeren10, brug12}.}
So far several dozens of clusters have been observed to have radio relics with a variety of morphologies
and most of them are considered to be associated with cluster merger activities
\citep[see for reviews, e.g.,][]{feretti12,brunetti2014}.
For instance, double radio relics, such as the ones in ZwCl0008.8+5215, are thought 
to reveal the bow shocks induced by a binary major merger \citep{vanweeren11b,gasperin14}. 
On the other hand, recently it was shown that shocks induced by the infall of the warm-hot
intergalactic medium (WHIM) along adjacent filaments into the hot intracluster medium (ICM) can efficiently 
accelerate CR electrons, and so they could be responsible for some radio relics in the cluster outskirts
\citep[see, e.g.,][]{hong14}.
{ The radio relic 1253+275 in Coma cluster observed in both radio \citep{brown11}
and X-ray \citep{ogreanbruggen13}
provides an example of such infall shocks.}

{
The so-called Sausage relic in CIZA J2242.8+5301 ($z=0.192$) contains a thin arc-like structure of $\sim 55$ kpc width 
and $\sim 2$~Mpc length, which could be represented by a portion of spherical
shell with radius $\sim 1.5$ Mpc \citep{vanweeren10}.
Unique features of this giant radio relic include 
the nearly uniform surface brightness along the length of the relic
and the strong polarization of up to $50-60\%$ with magnetic field vectors aligned with the relic \citep{vanweeren10}.
A temperature jump across the relic that corresponds to a $M_s\approx 2.54-3.15$ shock has been 
detected in X-ray observations \citep{akamatsu13, ogrean14}. This was smaller than $M_s\approx 4.6$
estimated from the above radio observation.
Several examples of Mpc-scale radio relics include
the Toothbrush relic in 1RXS J0603.3 with a peculiar linear morphology \citep{vanweeren2012}
and the relics in A3667 \citep{rottgering97} and A3376 \citep{bagchi06}.
The shock Mach numbers of radio relics estimated based on X-ray observation are often lower than
those inferred from the radio spectral index using the DSA model, for instance, in the Toothbrush relic \citep{ogrean13} and in the radio relic in A2256 \citep{trasatti15}.
Although such giant radio relics are quite rare, the fraction of X-ray luminous clusters 
hosting some radio relics is estimated to be $\sim 10$ \% or so \citep{feretti12}.
}

Through a number of studies using cosmological hydrodynamical simulations, it has been demonstrated
that during the process of hierarchical structure formation, abundant shocks are produced in the
large-scale structure of the universe, especially in clusters
\citep[e.g.,][]{ryu03,pfrommer2006,skillman2008,hoeft08,vazza09,vazza11,hong14}.
Considering that the characteristic time-scale of cluster dynamics including mergers is
$t_{\rm dyn}\sim 1$ Gyr, typical cluster shocks are expected to last for about the same period.
{ 
Yet, the number of observed radio relics, which is thought to trace such shocks, is still limited.}
So it is plausible to conjecture that cluster shocks may `turn on' to emit synchrotron radiation only
for a fraction of their lifetime.
One feasible scenario is that a cluster shock lights up in radio when it sweeps up a fossil
cloud, \ie a magnetized ICM gas with fossil relativistic electrons left over from either
a radio jet from AGN or a previous episode of shock/turbulence acceleration
(see the discussions in Section 2.5 and Figure 1).

Pre-exiting seed electrons and/or enhanced magnetic fields are the requisite conditions 
for possible lighting-up of relic shocks.
{
In particular, the elongated shape with {\it uniform} surface brightness 
and high polarization fraction of radio emission in 
the Sausage relic, may be explained, if a Mpc-scale thermal gas cloud, 
containing fossil relativistic electrons and permeated with
regular magnetic field of a few to several $\mu$G, is adopted.
A more detailed description will be given later in Section 2.5.}
In this picture, fossil electrons are expected to be re-accelerated for less than 
cloud-crossing time ($< R_{\rm cloud}/u_s\sim 100 \Myr$),
which is much shorter than the cluster dynamical time-scale.
In addition, only occasional encounters with fossil clouds combined with the short acceleration 
duration could alleviate the strong constraints on the DSA theory based on non-detection of 
$\gamma$-ray emission from clusters by Fermi-LAT \citep{ackermann14,vazza2015}.
A similar idea has been brought up by \citet{shimwell15}, who reported the discovery of a Mpc-scale,
elongated relic in the Bullet cluster 1E 0657-55.8.
They also proposed that the arc-like shape of uniform surface brightness in some radio
relics may trace the underlying regions of pre-existing, seed electrons remaining from old radio lobes.
{ On the other hand, \citet{ensslin01} suggested that radio relics could be explained by
revival of fossil radio plasma by compression due to a passage of a shock, rather than
DSA. In a follow-up study, \citet{ensslin02} showed using MHD simulations
that a cocoon of hot radio plasma swept by a shock turns into a filamentary or toroidal structure.
Although this scenario remains to be a viable explanation for some radio relics, 
it may not account for the uniform arc-like morphology of the Sausage relic.

It is now well established, through observations of radio halos/relics and Faraday rotation measures of background
radio sources,
that the ICM is permeated with $\mu$G-level magnetic fields \citep[e.g.][]{bonafede11,feretti12}.
The observed radial profile of magnetic field strength tends to peak at the center with a few $\mu$G and
decrease outward to $\sim 0.1 \mu$G in the cluster outskirts \citep{bonafede10}.
A variety of physical processes that could generate and amplify magnetic fields in the ICM have been suggested: 
primordial processes, plasma processes at the recombination epoch, and Biermann battery mechanism, combined with turbulence dynamo, 
in addition to galactic winds and AGN jets \citep[e.g.][]{ryu2008,dolag08,brug12,ryu2012,cho14}.
Given the fact that $\sim 5\mu$G fields are required to explain the amplitude and width of the observed radio flux profile of
the Sausage relic \citep[e.g.][]{vanweeren10,kang12}, 
the presence of a cloud with enhanced magnetic fields of several $\mu$G might be preferred to
the background fields of  $ \sim 0.1 \mu$G in the cluster periphery.
Alternatively, \cite{iapichino12} showed that the postshock magnetic fields can be amplified to 
$\sim 5-7\mu$G level leading to high degrees of polarization, if there exists dynamically significant turbulence  
in the upstream region of a curved shock. 
Although it is well accepted that magnetic fields can be amplified via various plasma instabilities at 
collisionless shocks, the dependence on the shock parameters such as the shock sonic and Alfv\'enic Mach 
numbers, and the obliquity of background 
magnetic fields remains to be further investigated \citep[see][]{schure12}. For example, 
the acceleration of protons and ensuing
magnetic field amplification via resonant and non-resonant streaming instabilities are found to be ineffective 
at perpendicular shocks \citep{bell78, bell04, capri14}.
}

In several studies using cosmological hydrodynamical simulations, synthetic radio maps
of simulated clusters were constructed by identifying shocks and adopting models 
for DSA of electrons and magnetic field amplification \citep{nuza12, vazza2012a, skillman2013, hong15}.
{ 
In particular, \citet{vazza12b} demonstrated, by generating mock radio maps of simulated cluster samples,
that radio emission tends to
increase toward the cluster periphery and peak around $0.2-0.5 R_{\rm vir}$ (where $R_{\rm vir}$ is the
virial radius), 
mainly because the kinetic energy dissipated at shocks peaks around $0.2 R_{\rm vir}$.
As a result, radio relics are rarely found in the cluster central regions.}
Re-acceleration of fossil relativistic electrons by cosmological shocks during the large scale structure
formation has been explored by \citet{pinzke13}.
The radio emitting shocks in these studies look like segments of spherical shocks, 
moving from the cluster core region into the periphery.
We presume that they are generated mostly as a consequence of major mergers or energetic
infalls of the WHIM along adjacent filaments.
So it seems necessary to study spherical shocks propagating through the cluster periphery,
rather than interpreting the radio spectra by DSA at {\it steady planar} shocks,
in order to better understand the nature of radio relics \citep{kang15a, kang15b}.

{
According to the DSA theory, in the case of a {\it steady planar} shock with {\it constant} postshock
magnetic field, the electron distribution function at the shock location becomes a power-law
of $f_e(p,r_s)\propto p^{-q}$, and so
the synchrotron emissivity from those electrons becomes a power-law of
$j_{\nu}(r_s)\propto \nu^{-\alpha_{\rm inj}}$.
The power-low slopes depend only on the shock sonic Mach number, $M_s$, and
are given as $q=4M_s^2/(M_s^2-1)$ and $\alpha_{\rm inj} = (M_s^2+3)/2(M_s^2-1)$
for the gasdynamic shock with the adiabatic index $\gamma_g=5/3$ \citep{dru83,be87,ensslin98}.
Here we refer $\alpha_{\rm inj}$ as the injection spectral index for a steady planar shock with constant postshock magnetic field.
Then, the volume-integrated synchrotron spectrum downstream
of the shock also becomes a simple power-law of $J_{\nu}= \int j_{\nu}(r)dV  \propto \nu^{-A_{\nu}}$
with the spectral index $A_\nu=\alpha_{\rm inj}+0.5$ above the break frequency, $\nu_{\rm br}$, 
since electrons cool via synchrotron and inverse-Compton (IC) losses behind the shock 
\citep[e.g.,][]{ensslin98,kang11}.}\footnote{Note that radio observers commonly use `$\alpha$' as the spectral index of the flux density, 
$S_\nu \propto \nu^{-\alpha}$ for unresolved sources, so in that case  $\alpha$ is the same as $A_\nu$. Here $\alpha_\nu(r)$ is defined as the spectral index of the {\it local} emissivity,
$j_\nu (r)$. See Equations (\ref{alpha})-(\ref{Anu}). }
Such predictions of the DSA theory have been applied to explain the observed properties of radio relics,
e.g., the relation between the injection spectral index and the 
volume-integrated spectral index, and the gradual steepening of
spatially resolved spectrum downstream of the shock.

{
\citet{kang12} performed time-dependent, DSA simulations of CR electrons for {\it steady planar} shocks 
with $M_s=2-4.5$ and constant postshock magnetic fields. Several models with thermal leakage injection or
pre-existing electrons were considered in order to reproduce the surface brightness and spectral 
aging profiles of radio relics in CIZA J2242.8+5301 and ZwCl0008.8+5215.
Adopting the same geometrical structure of radio-emitting volume as described in Section 2.5, they showed
that the synchrotron emission from shock accelerated electrons could explain the observed profiles 
of the radio flux, $S_\nu(R)$, of the Sausage relic, and the observed profiles of 
both  $S_\nu(R)$ and $\alpha_\nu(R)$ of the relic in ZwCl0008.8+5215.
Here $R$ is the distance behind the projected shock edge in the plane of the sky.
}

{
In the case of spherically expanding shocks with varying speeds and/or nonuniform magnetic field profiles,
on the other hand, the electron spectrum and the ensuing radio spectrum could deviate from those simple 
power-law forms, as shown in \citet{kang15a,kang15b}.
Then even the injection slope should vary with the frequency, \ie $\alpha_{\rm inj}(\nu)$.
Here we follow the evolution of a spherical shock expanding outward in the cluster outskirts with
a decreasing density profile, which may lead to a curvature in both the injected spectrum and the volume-integrated spectrum.
Moreover, if the shock is relatively young or the electron
acceleration duration is short ($\la 100$ Myr), then
the break frequency falls in $\nu_{\rm br} \sim 1$ GHz and 
the volume-integrated spectrum of a radio relic would steepen gradually with the spectral index from
$\alpha_{\rm inj}$ to $\alpha_{\rm inj}+0.5$ over $(0.1-10) \nu_{\rm br}$ \citep[e.g.][]{kang15b}.
}

In the case of the Sausage relic, \citet{vanweeren10} and \citet{stroe13} originally
reported observations of $\alpha_{\rm inj}\approx 0.6$ and $\alpha =A_\nu \approx 1.06$, which imply
a shock of $M_s\approx 4.6$.
\citet{stroe14b}, however, found a spectral steepening of the volume-integrated spectrum
at 16 GHz, which would be inconsistent with the DSA model for a steady planar shock.
Moreover, \citet{stroe14a}, by performing a spatially-resolved spectral fitting, revised
the injection index to a steeper value, $\alpha_{\rm inj}\approx 0.77$.
Then, the corresponding shock Mach number is reduced to $M_s\approx 2.9$.
They also suggested that the spectral age, calculated under the assumption of freely-aging
electrons downstream of a steady planar shock, might not be compatible with the shock speed 
estimated from X-ray and radio observations.
Also \citet{trasatti15} reported that for the relic in A2256,
the volume-integrated index steepens from $A_\nu \approx 0.85$ 
for $\nu = 351 - 1369$ MHz to $A_\nu \approx 1.0$ for $\nu = 1.37 - 10.45$ GHz,
which was interpreted as a broken power-law.

Discoveries of radio relic shocks with $M_s\sim 2-3$ in recent years have brought up the need for 
more accurate understanding of injections of protons and electrons at weak collisionless shocks,
especially at high plasma beta ($\beta_p\sim 50 - 100$) ICM plasmas \citep[e.g.][]{kang14}.
{ Here $\beta_p$ is the ratio of the gas to magnetic field pressure.}
{ 
Injection of electrons into the Fermi 1st-order process has been one of long-standing problems 
in the DSA theory for astrophysical shocks, because it involves complex plasma kinetic processes 
that can be
studied only through full Particle-in-Cell (PIC) simulations \citep[e.g.][]{amano09,riqu11}.}
It is thought that electrons must be pre-accelerated from their thermal momentum 
to several times the postshock thermal proton momentum to take part in the DSA process, 
and electron injection is much less efficient than proton injection due to smaller rigidity of electrons.
Several recent studies using PIC simulations have shown that some of incoming protons and electrons 
gain energies via shock drift acceleration (SDA) while drifting along the shock surface, and
then the particles are reflected toward the upstream region.
Those reflected particles can be scattered back to the shock
by plasma waves excited in the foreshock region, and then undergo multiple cycles of SDA,
resulting in power-law suprathermal populations \citep[e.g.,][]{guo14a, guo14b, park15}. 
Such `self pre-acceleration' of thermal electrons in the foreshock region could be sufficient
enough even at weak shocks in high beta ICM plasmas to explain the observed flux level of radio relics.
{ 
In these PIC simulations, however, subsequent acceleration of suprathermal electrons into 
full DSA regime has not been explored yet, because extreme computational resources 
are required to follow the simulations for a large dynamic range of particle energy.}

{ 
The main reasons that we implement the fossil electron distribution, instead
of the shock injection only case, are 
(1) the relative scarcity of radio relics compared to the abundance of
shocks expected to form in the ICM, (2) the peculiar uniformity of the surface brightness of the
Sausage relic, and (3) curved integrated spectra often found in some radio relics, implying
the acceleration duration $\la 100$~Myr, much shorter than the cluster dynamical time.}

{ 
In this paper, we consider a DSA model for radio relics;
a spherical shock moves into a magnetized gas cloud containing fossil relativistic electrons, 
while propagating through a density gradient in the cluster outskirts.
Specifically, we perform time-dependent DSA simulations for several spherical shock models with
the parameters relevant for the Sausage relic.
We then calculate the surface brightness profile, $I_{\nu}$, and the volume-integrated
radio spectrum, $J_{\nu}$, by adopting a specific geometrical structure of shock surface,
and compare them with the observational data of the Sausage relic.
}

In Section 2, the DSA simulations and the model parameters are described.
The comparison of our results with observations is discussed in Section 3.
A brief summary is given in Section 4.

\section{DSA SIMULATIONS OF CR ELECTRONS}

\subsection{1D Spherical CRASH Code}

{
We follow the evolution of the CR electron population by solving the following 
diffusion-convection equation in the one-dimensional (1D) spherical geometry:
\begin{eqnarray}
{\partial g_e\over \partial t}  + u {\partial g_e \over \partial r}
= {1\over{3r^2}} {{\partial (r^2 u) }\over \partial r} \left( {\partial g_e\over
\partial y} -4g_e \right)  
+ {1 \over r^2}{\partial \over \partial r} \left[r^2 D(r,p)  
{\partial g_e \over \partial r} \right]
+ p {\partial \over {\partial y}} \left( {b\over p^2} g_e \right),
\label{diffcon}
\end{eqnarray}
where $g_e(r,p,t)=f_e(r,p,t) p^4$ is the pitch-angle-averaged phase space distribution function
of electrons,
$u(r,t)$ is the flow velocity and $y \equiv \ln(p/m_e c)$ with the electron mass $m_e$ and
the speed of light $c$ \citep{skill75}.
The spatial diffusion coefficient, $D(r,p)$, is assumed to have 
a Bohm-like dependence on the rigidity,
\begin{equation}
D(r,p) = 1.7\times 10^{19} {\rm cm^2s^{-1}} \left({ B(r)\over 1\muG}\right)^{-1} 
\left({p \over m_e c}\right).
\label{Bohm}
\end{equation}
}
The cooling coefficient $b(p)=-dp/dt$ accounts for radiative cooling, and
the cooling time scale is defined as
\begin{equation}
 t_{\rm rad} (\gamma_e)= {p \over b(p)} =
9.8\times 10^{7} \yr \left({B_{\rm e} \over {5 \muG}}\right)^{-2} 
\left({\gamma_e \over 10^4 }\right)^{-1},
\label{trad}
\end{equation}
where $\gamma_e$ is the Lorentz factor of electrons.
Here the `effective' magnetic field strength, $B_{\rm e}^2 \equiv B^2 + B_{\rm rad}^2$ with
$B_{\rm rad}=3.24\muG(1+z)^2$, takes account for the IC loss due to the cosmic background radiation 
as well as synchrotron loss.
The redshift of the cluster CIZA J2242.8+5301 is $z=0.192$.

Assuming that the test-particle limit is applied at weak cluster shocks with $M_s \la$ several
(see Table 1), the usual gasdynamic conservation equations are solved to follow
the background flow speed, $u(r,t)$, using the 1D spherical version of
the CRASH (Cosmic-Ray Amr SHock) code \citep{kj06}.
The structure and evolution of $u(r,t)$ are fed 
into Equation (\ref{diffcon}), while the gas pressure, $P_g(r,t)$,
is used in modeling the postshock magnetic field profile (see Section 2.3).
We do not consider the acceleration of CR protons in this study,
since the synchrotron emission from CR electrons is our main interest and 
the dynamical feedback from the CR proton pressure can be ignored at weak shocks ($M_s\la 3$)
in the test-particle regime \citep{kang13}.
In order to optimize the shock tracking of the CRASH code, 
a comoving frame that expands with the instantaneous shock speed is adopted.
Further details of DSA simulations can be found in \citet{kang15a}.

\subsection{Shock Parameters}

{ 
To set up the initial shock structure, we adopt a Sedov self-similar blast wave 
propagating into a {\it uniform static} medium, which can be specified by two parameters, 
typically, the explosion energy, $E_0$, and the background density, $\rho_0$ \citep{ostriker88, ryu91}.
For our DSA simulations, we choose the initial shock radius and speed, $r_{s,i}$ and $u_{s,i}$,
respectively, and adopt the self-similar profiles of the gas density $\rho(r)$, the gas pressure $P_g(r)$,
and the flow speed $u(r)$ behind the shock in the upstream rest-frame.
For the fiducial case ({\bf SA1} model in Table 1), for example, the initial shock parameters are
$r_{s,i}=1.3~{\rm Mpc}$, and $u_{s,i}=3.3 \times 10^3\kms$.}
For the model parameters for the shock and upstream conditions, refer to Table 1 and Section 3.1.

We suppose that at the onset of the DSA simulations, this initial shock propagates 
through the ICM with the gas density gradient described by a power law of $r$.
{ 
Typical X-ray brightness profiles of observed clusters can be represented approximately
by the so-call beta model for isothermal ICMs,
$\rho(r)\propto [ 1+ (r/r_c)^2 ]^{-3\beta/2}$ with $\beta \sim 2/3$ \citep{sarazin86}.
In the outskirts of clusters, well outside of the core radius ($r\gg r_c$), it asymptotes as
$\rho(r)\propto r^{-3\beta}$. }
We take the upstream gas density of $\rho_{\rm up} \propto r^{-2}$ as the fiducial case ({\bf SA1}), 
but also consider $\rho_{\rm up}\propto r^{-4}$ ({\bf SA3}) and $\rho_{\rm up}=$ constant 
({\bf SA4}) for comparison.
The shock speed and Mach number decrease in time as the spherical shock expands, 
depending on the upstream density profile.
{
\citet{kang15b} demonstrated that
the shock decelerates approximately as $u_s\propto t^{-3/5}$ for $\rho_{\rm up}=$ constant
and as $u_s\propto t^{-1/3}$ for  $\rho_{\rm up}\propto r^{-2}$,  
while the shock speed is almost constant in the case of $\rho_{\rm up}\propto r^{-4}$. 
As a result, nonlinear deviations from the DSA predictions for steady planar shocks are expected 
to become the strongest in {\bf SA4} model, while the weakest in {\bf SA3} model.
In the fiducial {\bf SA1} model, which is the most realistic among the three models,
the effects of the evolving spherical shock are expected to be moderate.}  

{ 
The ICM temperature is set as $kT_1=3.35$keV, adopted from \citet{ogrean14}, in most of the models.
Hereafter, the subscripts ``1'' and ``2'' are used to indicate the quantities immediately upstream 
and downstream of the shock, respectively.
Although the ICM temperature is known to decrease slightly in the cluster outskirt, 
it is assumed to be isothermal, since the shock typically travels only $0.2-0.3$~Mpc for
the duration of our simulations $\la 100$~Myr.}

\subsection{Models for Magnetic Fields}

Magnetic fields in the downstream region of the shock are the key ingredient that governs the synchrotron
cooling and emission of CR electrons in our models.
We assume that the fossil cloud is magnetized to $\mu$G level.
As discussed in the Introduction, observations indicate that the magnetic field strength decreases from $\sim 1-10 \muG$ in
the core region to $\sim 0.1-1 \muG$ in the periphery of clusters \citep[e.g.,][]{bonafede10,feretti12}.
This corresponds to the plasma beta of
$\beta_p\sim 50 - 100$ in typical ICMs \citep[e.g.,][]{ryu2008,ryu2012}.
On the other hand, it is well established that magnetic fields can be amplified via resonant 
and non-resonant instabilities
induced by CR protons streaming upstream of {\it strong} shocks \citep{bell78,bell04}.
In addition, magnetic fields can be amplified by turbulent motions behind shocks \citep{giacal07}.
Recent hybrid plasma simulations have shown that the magnetic field amplification
factor due to streaming CR protons scales with the Alfv\'enic Mach number, $M_A$, 
and the CR proton acceleration efficiency 
as $\langle \delta B/B\rangle^2 \sim$ $3 M_A (P_{\rm cr,2}/\rho_1 u_s^2)$ \citep{capri14}.
Here, $\delta B$ is the turbulent magnetic field perpendicular to the mean background magnetic field,
$P_{\rm cr,2}$ is the downstream CR pressure, and
$\rho_1 u_s^2$ is the upstream ram pressure.
For typical radio relic shocks, the sonic and Alfv\'enic Mach numbers are expected to range
$2 \lesssim M_s \lesssim 5$ and $10 \lesssim M_A \lesssim 25$, respectively \citep[e.g.,][]{hong14}.

{ 
The magnetic field amplification in both the upstream and downstream of {\it weak} shocks is 
not yet fully understood, especially in high beta ICM plasmas.
So we consider simple models for the postshock magnetic fields.
For the fiducial case, we assume that the magnetic field strength across the shock transition is increased 
by compression of 
the two perpendicular components:
\begin{equation}
B_2(t)=B_1 \sqrt{1/3+2\sigma(t)^2/3}, 
\label{b2}
\end{equation}
where $B_1$ and $B_2$ are the magnetic field strengths immediately upstream and downstream of the shock,
respectively, and $\sigma(t)=\rho_2/\rho_1$ is the time-varying compression ratio across the shock.
For the downstream region ($r<r_s$), the magnetic field strength is assumed to scale with the 
gas pressure:
\begin{equation}
B_{\rm dn}(r,t)= B_2(t) \cdot [P_g(r,t)/P_{g,2}(t)]^{1/2},
\label{bd}
\end{equation}
where $P_{g,2}(t)$ is the gas pressure immediately behind the shock.
This assumes that the ratio of the magnetic to thermal energy density is constant downstream of the shock.
Since $P_g(r)$ decreases behind the spherical blast wave, $B_{\rm dn}(r)$ also decreases downstream as
illustrated in Figure 2. 
This fiducial magnetic field model is adopted in most of the models described in Table 1, except {\bf SA2} and {\bf SA4} models.
The range of $B_2(t)$ is shown for the acceleration duration of $0 \leq t_{\rm age} \leq 60$ Myr
in Table 1, reflecting the decrease
of shock compression ratio during the period, but the change is small.

In the second, more simplified (but somewhat unrealistic) model, 
it is assumed that $B_1= 2\muG$ and $B_2=7\muG$, 
and the downstream magnetic field strength is constant, \ie, $B_{\rm dn} = B_2$ for $r<r_s$. 
This model was adopted in \citet{kang12} and also for {\bf SA2} and {\bf SA4} models 
for comparison in this study.

\citet{kang15b} showed that the postshock synchrotron emission increases downstream away from the shock
in the case of a decelerating shock, because the shock is stronger at earlier time.
But such nonlinear signatures become less distinct in the model with decreasing downstream 
magnetic fields, compared with the model with constant downstream magnetic fields,
because the contribution of synchrotron emission from further downstream region becomes weaker.}

\subsection{Injection Momentum}

As described in the Introduction, the new picture of particle injection emerged from recent PIC simulations
is quite different from the `classical' 
thermal leakage injection model commonly employed in previous studies of DSA \citep[e.g.,][]{kjg02}.
{ 
However, the requirement of $p\gtrsim 3 p_{\rm th,p}$ for particles to take part in the full Fermi 1st-order process,
scattering back and forth diffusively across the shock transition zone with thickness,
$\Delta l_{\rm shock} \sim r_g(p_{\rm th,p})$,
seems to remain valid \citep{capriolietal15, park15}.
Here, $p_{\rm th,p}=\sqrt{2m_p k_B T_2}$ is the most probable momentum of
thermal protons with postshock temperature $T_2$ and $r_g$ is the gyroradius of particles.}
In other words, only suprathermal particles with the gyro-radius greater than the shock thickness are 
expected to cross the shock transition layer.
Hence, we adopt the traditional phenomenological model in which only particles above the injection momentum,
$p_{\rm inj} \approx 5.3\ m_p u_s/\sigma$, are allowed to get injected into the CR populations 
at the lowest momentum boundary \citep{kjg02}.
{
This can be translated to the electron Lorentz factor,
$\gamma_{\rm e,inj}=p_{\rm inj}/m_e c \sim 30 (u_s/3000\kms) (3.0/\sigma)$.}
In the case of expanding shocks considered in this study, $p_{\rm inj}(t)$ decreases 
as the shock slows down in time. 

\subsection{Fossil Electrons}

As mentioned in the Introduction, one peculiar feature of the Sausage relic is the {\it uniform} surface brightness along the Mpc-scale
arc-like shape, which requires a special geometrical distribution of shock-accelerated electrons
\citep{vanweeren11a}.
Some of previous studies adopted the ribbon-like curved shock surface and the downstream swept-up
volume, viewed edge-on with the viewing extension angle $\psi\sim 10^{\circ}$ 
\citep[e.g.,][]{vanweeren10,kang12,kang15b}.
{ 
We suggest a picture where a spherical shock of radius $r_s \sim 1.5$ Mpc passes through
an elongated cloud with width $w_{\rm cloud}\sim 260$~kpc and length $l_{\rm cloud}\sim 2$~Mpc,
filled with fossil electrons.
Then the shock surface penetrated into the cloud becomes a ribbon-like patch, distributed on a sphere with radius
$r_s \sim 1.5$ Mpc with the angle $\psi=360^{\circ} \cdot w_{\rm cloud}/(2\pi r_s) \sim 10^{\circ}$.
The downstream volume of radio-emitting, reaccelerated electrons has the width
$\Delta l(\gamma_e) \approx (u_s/\sigma) \cdot {\rm min}[t_{\rm age},t_{\rm rad}(\gamma_e)]$, as shown in 
Figure 1 of \citet{kang15b}.
Hereafter the `acceleration age', $t_{\rm age}$, is defined as the duration of electron acceleration 
since the shock encounters the cloud.
This model is expected to produce a uniform surface brightness along the relic length.
Moreover, if the acceleration age is $t_{\rm age} \lesssim 100 \Myr$, 
the volume-integrated radio spectrum is expected to steepen gradually over 0.1-10~GHz \citep{kang15b}.
}

{ 
There are several possible origins for such clouds of relativistic electrons in the ICMs:
(1) old remnants of radio jets from AGNs, (2) electron populations that were accelerated by 
previous shocks and have cooled down below $\gamma_e < 10^4$, and
(3) electron populations that were accelerated by turbulence during merger activities.
Although an AGN jet would have a hollow, cocoon-like shape initially, it may turn into a
filled, cylindrical shape through diffusion, turbulent mixing, or contraction of relativistic 
plasmas, as the electrons cool radiatively.
During such evolution relativistic electrons could be mixed with the ICM gas within the cloud. 
We assume that the cloud is composed of the thermal gas of $\gamma_g=5/3$, 
whose properties (density and temperature) are the same as the surrounding ICM gas, 
and an additional population of fossil electrons with dynamically insignificant pressure.
In that regard, our fossil electron cloud is different from hot bubbles considered in previous
studies of the interaction of a shock with a hot rarefied bubble
\citep[e.g.,][]{ensslin02}.
In the other two cases where electrons were accelerated either by shocks or by turbulence
\citep[see, e.g.,][]{farns2013}, it is natural to assume that
the cloud medium should contain both thermal gas and fossil electrons.
}

{
Three different spectra for fossil electron populations are considered.
In the first fiducial case (\eg {\bf SA1} model), nonthermal electrons have the momentum around $p_{\rm inj}$,
which corresponds to $\gamma_{\rm e,inj} \sim 20-30$ for the model shock parameters considered here.
So in this model, seed electrons with $\sim \gamma_{\rm e,inj}$ 
are injected from the fossil population and re-accelerated into radio-emitting CR electrons with
$ \gamma_e \ga 10^3$.}
Since we compare the surface brightness profiles in arbitrary units here, 
we do not concern about the normalization of the fossil population.

In the second model, the fossil electrons are assumed to have a power-law distribution extending 
up to $\gamma_e \gg 10^4$,
\begin{equation}
f_{\rm e,up}(p)=f_0\cdot \left(p \over p_{\rm inj}\right)^{-s}.
\label{f1}
\end{equation}
For the modeling of the Sausage relic, the value of $s$ is chosen as $s= 2\alpha_{\rm inj}+3=4.2$
with $\alpha_{\rm inj}=0.6$ ({\bf SA1p} model).

As mentioned in the Introduction, the volume-integrated radio spectrum of the Sausage relic 
seems to steepen at high frequencies, perhaps more strongly than expected 
from radiative cooling alone \citep{stroe14b}.
So in the third model, we consider a power-law population with exponential cutoff as follows: 
\begin{equation}
f_{\rm e,up}(p)=f_0\cdot \left(p \over p_{\rm inj}\right)^{-s} \exp \left[ - \left({\gamma_e \over \gamma_{e,cut}} \right)^2 \right],
\label{f1exp}
\end{equation}
where $\gamma_{e,cut}$ is the cutoff Lorentz factor.
This may represent fossil electrons that have cooled down to $\sim \gamma_{e,cut}$
from the power-law distribution in Equation (\ref{f1}). 
The integrated spectrum of the Sausage relic shows a strong curvature above $\sim 1.5$ GHz \citep{stroe14b},
which corresponds to the characteristic frequency of synchrotron emission from electrons with
$\gamma_e \sim 1.5\times 10^4$ when the magnetic field strength ranges $5-7 \muG$
(see Equation (\ref{nupeak}) below).
So $\gamma_{e,cut} = 10^4$ and $2 \times 10^4$ are chosen for {\bf SC1pex1} and {\bf SC1pex2} models, respectively.

{Figure 3 shows the electron distributions in the cloud medium: the thermal distribution 
for the ICM gas with $kT = 3.35$ keV and the fossil population.
The characteristics of the different models, {\bf SA1} (fiducial model),
{\bf SA1p} (power-law), and {\bf SC1pex1} (power-law with an exponential cutoff) will be described in Section 3.1.
Note that there could be `suprathermal' distribution between the thermal and CR populations.
But it does not need to be specified here, since only $f(p)$ around $p_{\rm inj}$ controls the
spectrum of re-accelerated electrons.}

\section{RESULTS OF DSA SIMULATIONS}

\subsection{Models}

We consider several models whose characteristics are summarized in Table 1.
{ 
For the fiducial model, {\bf SA1}, 
the upstream density is assumed to decrease as $\rho_{\rm up}=\rho_0(r/r_{s,i})^{-2}$, while 
the upstream temperature is taken to be $kT_1 = 3.35$ keV.
Since we do not concern about the absolute radio flux level in this study, 
$\rho_0$ needs not to be specified.
The preshock magnetic field strength is $B_1=2.5\muG$ and the immediate postshock 
magnetic field strength is $B_2(t)\approx 6.7-6.3\muG$ during 60 Myr, while 
the downstream field, $B_{\rm dn}(r)$, is given as in equation (\ref{bd}).
The initial shock speed is $u_{s,i}=3.3\times 10^3\kms$, corresponding to the sonic Mach number
of $M_{s,i}=3.5$ at the onset of simulation.
In all models with $\rho_{\rm up} \propto r^{-2}$, the shock slows down as $u_s\propto t^{-1/3}$,
and at the electron acceleration age of 60 Myr, $u_s\approx 2.9\times 10^3\kms$ with $M_s\approx 3.1$ and $\alpha_{\rm inj}\approx 0.73$.
The fossil seeds electrons are assumed to have a delta-function-like distribution around $\gamma_{\rm e,inj} \approx 30$.

In {\bf SA1b} model, the upstream magnetic field strength is weaker with $B_1=0.25 \muG$
and $B_2(t)\approx 0.67-0.63\muG$ during 60 Myr. Otherwise, it is the same as the fiducial model.
So the character `b' in the model name denotes `weaker magnetic field', compared to the fiducial model.
Comparison of {\bf SA1} and {\bf SA1b} models will be discussed in Section 3.4.

In {\bf SA1p} model, the fossil electrons have a power-law population of $f_{\rm e,up} \propto p^{-4.2}$,
while the rest of the parameters are the same as those of {\bf SA1} model.
The character `p' in the model name denotes a `power-law' fossil population.

In {\bf SA2} model, both the preshock and postshock magnetic field strengths are constant, \ie
$B_1=2\muG$ and $B_2=7\muG$, otherwise it is the same as {\bf SA1} model.

In {\bf SA3} model, the upstream gas density decreases as $\rho_{\rm up}=\rho_0(r/r_{s,i})^{-4}$, so the
shock decelerates more slowly, compared to {\bf SA1} model.
Considering this, the initial shock speed is set to $u_{s,i}=3.0\times 10^3\kms$ with $M_{s,i}=3.2$.
At the acceleration age of 60 Myr, the shock speed decreases to $u_s\approx 2.8 \times 10^3\kms$ 
corresponding to $M_s\approx 3.0$.

In {\bf SA4} model, the upstream density is constant, so the shock decelerates approximately as 
$u_s \propto t^{-3/5}$, more quickly, compared to {\bf SA1} model.
The upstream and downstream magnetic field strengths are also constant as in {\bf SA2} model.
Figure 2 compares the profiles of the flow speed, $u(r,t)$, and the magnetic field strength, $B(r,t)$,
in {\bf SA1} and {\bf SA4} models. 
Note that although the shocks in {\bf SA1} and {\bf SA4} models are not very strong with the initial Mach number $M_{s,i} \approx 3.5$,
they decelerate approximately as $u_s\propto t^{-1/3}$ and
$u_s\propto t^{-3/5}$, respectively, as in the self-similar solutions of blast waves
\citep{ostriker88,ryu91}.
The shock speed decreases by $\sim 12 - 15 \%$ during the electron acceleration age of 60 Myr in the two models.

In {\bf SB1} model, the preshock temperature, $T_1$, is lower by a factor of $1.5^2$,
and so the initial shock speed $u_{s,i}=3.3\times 10^3\kms$ corresponds to $M_{s,i}\approx 5.3$.
The shock speed $u_s\approx 2.9 \times 10^3\kms$ at the age of 60 Myr corresponds to $M_s\approx 4.6$, 
so the injection spectral index $\alpha_{\rm inj}\approx 0.6$.
The `{\bf SB}' shock is different from the `{\bf SA}' shock in terms of only the sonic Mach number.

In {\bf SC1pex1} and {\bf SC1pex2} models, the fossil electrons have
$f_{\rm e,up} \propto p^{-4.2} \cdot \exp[-(\gamma_e/\gamma_{e,cut})^2]$ with the cutoff at
$\gamma_{e,cut}=10^4$ and $\gamma_{e,cut}=2\times 10^4$, respectively.
The character `pex' in the model name denotes a `power-law with an exponential cutoff'.
A slower initial shock with $u_{s,i}\approx 2.2\times 10^3 \kms$ and $M_{s,i}\approx 2.4$ is chosen,
so at the acceleration age of 80 Myr the shock slows down to $M_s\approx 2.1$ with 
$\alpha_{\rm inj}\approx 1.1$.
The `{\bf SC}' shock differs from the `{\bf SA}' shock in terms of the shock speed and the sonic Mach number.}
The integrated spectral index at high frequencies would be steep with $A_{\nu }\approx 1.6$,
while $A_{\nu} \approx 0.7$ at low frequencies due to the flat fossil electron population.
They are intended to be toy models that could reproduce the integrated spectral indices, $A_{\nu} \sim 0.7$
for $\nu = 0.1 - 0.2$ GHz and $A_{\nu}\sim 1.6$ for $\nu = 2.3 - 16$ GHz, compatible with
the observed curved spectrum of the Sausage relic \citep{stroe14b}.

\subsection{Radio Spectra and Indices}

The local synchrotron emissivity, $j_{\nu}(r)$, is calculated, 
using the electron distribution function, $f_e(r,p,t)$, and the magnetic
field profile, $B(r,t)$. 
Then, the radio intensity or surface brightness, 
$I_{\nu}$, is calculated by integrating $j_{\nu}$ along lines-of-sight (LoSs).
\begin{equation}
I_{\nu}(R)= 2 \int_0^{h_{\rm max}} j_{\nu}(r) d {\it h}. 
\label{SB}
\end{equation}
$R$ is the distance behind the projected shock edge in the plane of the sky, as defined in the Introduction,
and $h$ is the path length along LOSs;
$r$, $R$, and $h$ are related as $r^2= (r_s-R)^2 + h^2$.
The extension angle is $\psi=10^{\circ}$ (see Section 2.5).
{
Note that the radio flux density, $S_{\nu}$, can be obtained by convolving $I_{\nu}$ with a telescope beam as
$S_{\nu}(R) \approx I_{\nu}(R) \pi \theta_1 \theta_2 (1+z)^{-3}$,
if the brightness distribution is broad compared to the beam size of $\theta_1 \theta_2$.}

The volume-integrated synchrotron spectrum, $J_{\nu}=\int j_{\nu}(r) dV $, 
is calculated by integrating $j_{\nu}$ over the entire downstream region with the assumed
geometric structure described in Section 2.5.
The spectral indices of the local emissivity, $j_{\nu}(r)$, and  the integrated spectrum, $J_{\nu}$, are defined as follows:
\begin{equation}
\alpha_{\nu}(r) =- {{d\ln j_{\nu}(r) }\over {d\ln \nu}},
\label{alpha}
\end{equation}
\begin{equation}
A_{\nu} =- {{d\ln J_{\nu} }\over {d\ln \nu}}. 
\label{Anu}
\end{equation}
{ 
As noted in the Introduction, unresolved radio observations usually report the spectral index, `$\alpha$' at various frequencies, which
is equivalent to $A_{\nu}$ here.}

In the postshock region, the cutoff of the electron spectrum, $f_e(r,\gamma_e)$,
decreases downstream from the shock due to radiative losses, 
so the volume-integrated electron spectrum steepens from 
$\gamma_e^{-q}$ to $\gamma_e^{-(q+1)}$ above the break Lorentz factor (see Figure 5).
At the electron acceleration age $t_{\rm age}$, the break Lorentz factor can be estimated from the condition 
$t_{\rm age}=t_{\rm rad}$ \citep{kang12}:
\begin{equation}
\gamma_{\rm e,br}  \approx   10^4 \left({t_{\rm age} \over 100 {\rm Myr}}\right)^{-1} 
\left( {5^2} \over {B_2^2+B_{\rm rad}^2}  \right).
\label{pbr}
\end{equation}
Hereafter, the postshock magnetic field strength, $B_2$, and $B_{\rm rad}$ are expressed in units of $\mu$G.
Since the synchrotron emission from mono-energetic electrons with $\gamma_{\rm e}$ peaks around
\begin{equation}
\nu_{\rm peak} \approx 0.3 \left({3eB_2 \over 4\pi{m_e c}}\right) \gamma_e^2 
\approx 0.63\ {\rm GHz} \cdot \left({B_2\over 5 }\right) \left({\gamma_e \over 10^4}\right)^2,
\label{nupeak}
\end{equation}
the break frequency that corresponds to $\gamma_{\rm e,br}$ becomes
\begin{equation}
\nu_{\rm br}\approx 0.63 {\rm GHz} \left( {t_{\rm age} \over {100 \rm Myr}} \right)^{-2}
 \left( {5^2} \over {B_2^2+B_{\rm rad}^2}  \right)^{2} \left( {B_{2} \over {5}} \right).
\label{fbr}
\end{equation}
{ 
So the volume-integrated synchrotron spectrum, $J_{\nu}$, has a spectral break,  or more precisely a 
{\it gradual} increase of the spectral index approximately from $\alpha_{\rm inj}$ to $\alpha_{\rm inj}+0.5$ around
$\nu_{\rm br}$.}

\subsection{Width of Radio Shocks}

For electrons with $\gamma_e>10^4$, the radiative cooling time in Equation (\ref{trad})
becomes shorter than the acceleration age if $t_{\rm age} \ga 100$ Myr. 
Then, the width of the spatial distribution of those high-energy electrons downstream of the shock
becomes
\begin{equation}
\Delta l_{\rm cool}(\gamma_e) \approx u_2 t_{\rm rad}(\gamma_e) 
\approx 100\ {\rm kpc} \cdot \left({ u_2 \over {10^3 \kms}}\right)
\left( { {5^2} \over {B_2^2+B_{\rm rad}^2}}\right)
 \left({\gamma_e \over 10^4}\right)^{-1},
\label{lcool}
\end{equation}
where $u_2$ is the downstream flow speed.

With the characteristic frequency $\nu_{\rm peak}$ of electrons with $\gamma_{\rm e}$ in Equation
(\ref{nupeak}), the width of the synchrotron emitting region behind the shock at the observation frequency
of $\nu_{\rm obs}= \nu_{\rm peak}/(1+z)$ would be similar to $\Delta l_{\rm cool}(\gamma_e)$:
\begin{equation}
\Delta l_{\nu_{\rm obs}} \approx W \cdot \Delta l_{\rm cool}(\gamma_e) 
\approx W\cdot 100\ {\rm kpc} \cdot \left({ u_2 \over {10^3 \kms}}\right)
\left( { {5^2} \over {B_2^2+B_{\rm rad}^2}}\right) \left({B_2 \over 5}\right)^{1/2}
\left[{{\nu_{\rm obs}(1+z) \over {0.63{\rm GHz}}} }\right]^{-1/2}.
\label{lwidth}
\end{equation}
Here, $W$ is a numerical factor of $\sim 1.2-1.3$.
This factor takes account for the fact that
the spatial distribution of synchrotron emission at $\nu_{\rm peak}$ is somewhat broader than
that of electrons with the corresponding $\gamma_e$, because more abundant, lower energy electrons
also make contributions \citep{kang15a}.
One can estimate two possible values of the postshock magnetic field strength, $B_2$, from this relation,
if $\Delta l_{\nu_{\rm obs}}$ can be determined from the observed profile of surface brightness and 
$u_2$ is known.
For example, \citet{vanweeren10} inferred two possible values of the postshock magnetic field strength 
of the Sausage relic, $B_{\rm low} \approx 1.2\muG$ and $B_{\rm high} \approx 5\muG$, by assuming that
the FWHM of surface brightness, $\Delta l_{\rm SB}$, is the same as
$\Delta l_{\nu_{\rm obs}} \approx \Delta l_{\rm cool}(\gamma_e)\approx 55$ kpc 
(i.e., $W=1$).

On the other hand, \citet{kang12} showed that the profile of surface brightness
depends strongly on $\psi$ in the case of the assumed geometrical structure.
Figure 4 illustrates the results of a {\it planar} shock with different $\psi$'s.
The shock has $u_s=2.7\times 10^3 \kms$ and $M_s=4.5$.
The magnetic field strengths are assumed to be $B_1=2\muG$ and $B_2=7\muG$, upstream and downstream of
the shock, respectively.
The results shown are at the acceleration age of 80 Myr.
{ 
Note that the quantities, $g_e(d,\gamma_e)$, $j_{\nu}(d)$, and  $I_{\nu}(R)$, are multiplied by arbitrary powers of $\gamma_e$ and $\nu_{\rm obs}$, so they can be shown for different values of
$\gamma_e$ and $\nu_{\rm obs}$ with single linear scales in Figure 4.}
Here, the FWHM of $g_e(\gamma_e)$ is, for instance, $\Delta l_{\rm cool}(\gamma_e)\approx 28$ kpc for
$\gamma_e=8.47\times 10^3 $ (red dotted line in the top panel),
while the FWHM of $j_{\nu}$ is $\Delta l_{\nu_{\rm obs}}\approx 35$ kpc for $\nu_{\rm obs}=0.594$ GHz 
(red dotted line in the middle panel).
Note that the values of $\gamma_e$ and $\nu_{\rm peak}= \nu_{\rm obs}(1+z)$ are chosen from the sets of 
discrete simulation bins, so they satisfy only approximately the Equation (\ref{nupeak}).
The bottom panel demonstrates that the FWHM of $I_{\nu}$, $\Delta l_{\rm SB}$, strongly depends
on $\psi$.
For instance, $\Delta l_{\rm SB} \approx 47$ kpc for $\nu_{\rm obs}=0.594$ GHz,
if $\psi=10^{\circ}$ (red dotted line).
This implies that due to the projection effect, $\Delta l_{\rm SB}$ could be substantially
different from $\Delta l_{\rm cool}$.
So $\Delta l_{\rm SB}$ of the observed radio flux profile 
may not be used to infer possible values of $B_2$ in radio
relics, unless the geometrical structure of the shock is known and so the projection effect can be modeled
accordingly.

Note that the quantity $Y(B_2,z)\equiv [({B_2^2+B_{\rm rad}^2})/ 5^2]^{-1} \cdot({B_2/ 5})^{1/2}$
in Equation (\ref{lwidth}) also appears  as $Y^2$ in Equation (\ref{fbr}).
So the break frequency becomes identical for the two values of $B_2$, $B_{\rm low}$ and $B_{\rm high}$, 
that give the same value of $Y$.
For example, {\bf SA1} model with $B_2(t)\approx 6.7-6.3 \muG$ and {\bf SA1b} model with 
$B_2(t)\approx 0.67-0.63 \muG$ (see Table 1) produce not only the similar $\Delta l_{\nu_{\rm obs}}$
but also the similar spectral break in $J_{\nu}$.
But the corresponding values of $\gamma_{\rm e,br}$ in Equation (\ref{pbr}) are different
for the two models with $B_{\rm low}$ and $B_{\rm high}$.
Moreover, the amplitude of the integrated spectrum would scale as
$J_{\nu}(B_{\rm high})/J_{\nu}(B_{\rm low}) \sim (B_{\rm high}/B_{\rm low})^2$,
if the electron spectrum $n_e(\gamma_e)$ for $\gamma_e < \gamma_{\rm e,br}$ is similar
for the two models .
We compare these two models in detail in the next section.

\subsection{Comparison of the Two Models with $B_{\rm high}$ and $B_{\rm low}$}

In Figure 5, we first compare the electron spectra and the synchrotron emission spectra
in {\bf SA1} and {\bf SA1b} models.
Here, the plotted quantities, $g_e$, $G_e$, $j_{\nu}$ and $J_{\nu}$, are in arbitrary units.
The postshock magnetic field strength decreases from $B_2\approx 6.7 \muG$ to $6.2\muG$ in 110 Myr
in {\bf SA1} model, while $B_2\approx 0.67 \muG$ to $0.62\muG$ in {\bf SA1b} model.
In these two models, the values of $Y(B_2,z)$ are similar,
so the spectral break in the integrated spectra should be similar as well.

The left panels of Figure 5 show that
the electron spectrum at the shock, $g_e(r_s,p)$, steepens in time as the shock weakens.
As expected, the volume-integrated electron spectrum, $G_e(p) = p^4 F_e(p) = p^4 \int f_e(r,p) dV$,
steepens by one power of $p$ above the break momentum, $p_{\rm e,br}$, which decreases in time 
due to radiative losses.
In both models, however, the slopes, $q=-d \ln f_e(r_s,p)/d \ln p$ and $Q=-d \ln F_e(p)/d \ln p$,
deviate from the simple steepening of $Q=q$ to $q+1$ above $p_{\rm e,br}$,
because of the time-dependence of the shock properties.

The right panels of Figure 5 show the synchrotron spectrum at the shock, $j_{\nu}(r_s)$,
the volume-integrated synchrotron spectrum, $J_{\nu}$, 
and their spectral indices, $\alpha_{\rm inj}(\nu)$ and $A_{\nu}$.
Note that in both models,
the transition of $A_{\nu}$ from $\alpha_{\rm inj}$ to $\alpha_{\rm inj}+0.5$ is broader
than the transition of $Q$ from $q$ to $q+1$.
This is because the synchrotron emission at a given frequency comes
from electrons with a somewhat broad range of $\gamma_e$'s.
As in the case of $Q$, $A_{\nu}$ does not follow the simple steepening, but
rather shows nonlinear features due to the evolving shock properties.
This implies that the simple relation, $A_{\nu}\approx \alpha_{\rm inj} + 0.5$,
which is commonly adopted in order to confirm the DSA origin of synchrotron spectrum,
should be applied with some caution in the case of evolving shocks.

The highest momentum for $g_e(r_s,p)$, 
$p_{\rm eq}\propto u_s \cdot [ B_1/ (B_{e,1}^2 + B_{e,2}^2) ]^{1/2}$, 
is higher in {\bf SA1} model with stronger magnetic field,
but the amplitude of $g_e(r_s,p)$ near $p_{\rm eq}$,
say at $p/m_e c\sim 10^6-10^{7.5}$, is greater in {\bf SA1b} model with weaker magnetic field.
As a consequence, the ratio of the synchrotron emission of the two models is somewhat less 
than the ratio of the magnetic energy density, $(B_{\rm high}/B_{\rm low})^2=100$.
In {\bf SA1b} model, for example, the amplitudes of $j_{\nu}(r_s)$ and $J_{\nu}$ at $0.1-10$ GHz 
are reduced by a factor of $\sim 60$, compared to the respective values in {\bf SA1} model.
Also the cutoff frequencies for both $j_{\nu}(r_s)$ and $J_{\nu}$ are lower in {\bf SA1b} model,
compared to those in {\bf SA1} model.
As pointed above, $\nu_{\rm br}$ is almost the same in the two models, although $p_{\rm e,br}$ 
is different.
This confirms that we would get two possible solutions for $B_2$, if we attempt to estimate
the magnetic field strength from the value of $\nu_{\rm br}$ in the integrated spectrum.

\subsection{Surface Brightness Profile}

{
As noted in Section 3.3,
the width of the downstream electron distribution behind the shock is determined by the advection length, 
$\Delta l_{\rm adv}\approx u_2 \cdot t_{\rm age}$, for low-energy electrons
or by the cooling length, $\Delta l_{\rm cool}(\gamma_e)\approx u_2 \cdot t_{\rm rad}(\gamma_e)$,
for high-energy electrons.
As a result, at high frequencies the width of the synchrotron emission region,
$\Delta l_{\nu_{\rm obs}}$, varies with the downstream magnetic field strength 
for given $u_2$ and $\nu_{\rm obs}$ as in Equation (\ref{lwidth}).
In addition, the surface brightness profile, $I_{\nu}(R)$, 
and its FWHM, $\Delta l_{\rm SB}$, also depend on the extension angle, 
$\psi$, as demonstrated in Figure 4.}

In Figures 6 and 7, the spatial profiles of $I_{\nu}(R)$
are shown for eight models listed in Table 1.
Here, we choose the observation frequencies $\nu_{\rm obs}=150,\ 600,\ 1400$ MHz;
the source frequencies are given as $\nu = \nu_{\rm obs}(1+z)$ with $z=0.192$.
While the downstream flow speed ranges $u_2\approx 1,000 - 900\kms$ in most of the models,
$u_2\approx 850-750 \kms$ in {\bf SC1pex1} model.
So the results are shown at $t_{\rm age} = 30,\ 60,$ and $110 \Myr$, except in {\bf SC1pex1} 
model for which the results are shown at $t_{\rm age} = 30,\ 80,$ and $126 \Myr$.
Note that the quantity, $\nu I_{\nu}\times X$, is shown in order to be plotted 
with one linear scale, where $X$ is the numerical scale factor specified in each panel.
In {\bf SB1} model with a stronger shock, for example,
the electron acceleration is more efficient by a factor of about $10$,
compared to other models, so the profiles shown are reduced by similar factors.

For a fixed value of $\psi$, the surface brightness profile is determined by the distribution of
$j_{\nu}(r)$ behind the shock along the path length $h$, as given in Equation (\ref{SB}).
The intensity increases gradually from the shock position ($R=0$) to the first {\it inflection point}, 
$R_{\rm inf,1}(t)=r_s(t)(1- \cos \psi)\approx 21-24$ kpc with $\psi=10^{\circ}$,
mainly due to the increase of the path length along LoSs.
Since the path length starts to decrease beyond $R_{\rm inf,1}$,
if the emissivity, $j_{\nu}$, is constant or decreases downstream of the shock,
then $I_{\nu}$ should decrease for $R>R_{\rm inf,1}$.
In all the models considered here, however, at low frequencies, 
$j_{\nu}$ increases downstream of the shock, 
because the model spherical shock is faster and stronger at earlier times.
As a result, $I_{\nu}$ at 150 MHz is almost constant or decrease very slowly beyond $R_{\rm inf,1}$,
or even increases downstream away from the shock in some cases (\eg {\bf SA2} and {\bf SA4} models).
So the downstream profile of the synchrotron radiation at low frequencies emitted by uncooled,
low-energy electrons could reveal some information about the shock dynamics, providing that the downstream 
magnetic field strength is known.

The second inflection in $I_{\nu}(R)$ occurs roughly at $R_{\rm inf,2}\approx W \cdot \Delta l_{\rm adv}$
for low frequencies, and at $R_{\rm inf,2} \approx W \cdot \Delta l_{\rm cool}(\gamma_e)$ for 
high frequencies where $t_{\rm rad} < t_{\rm age}$, with $W\approx 1.2-1.3$.
Here, $\Delta l_{\rm adv}\approx 100\ {\rm kpc} (u_2/10^3\kms)(t_{\rm age}/100\ {\rm Myr})$,
and $\Delta l_{\rm cool}(\gamma_e)$ is given in Equation (\ref{lcool}).
{
Figures 6 and 7 exhibit that at 30 Myr ($t_{\rm age} < t_{\rm rad}$), the second inflection appears
at the same position ($\Delta l_{\rm adv}\approx 30$~kpc) for the three frequencies shown.
At later times, the position of the second inflection depends on $t_{\rm age}$ at low frequencies,
while it varies with $B_2$ and $\nu_{\rm obs}$ in addition to $u_2$.
Thus, only if $t_{\rm age} > t_{\rm rad}(\gamma_e)$, $\Delta l_{\rm SB}$ at high frequencies can
be used to infer $B_2$, providing that $\psi$ and $u_2$ are known.
}

The width of the Sausage relic, defined as the FWHM of the surface brightness, was estimated to be
$\Delta l_{\rm SB} \sim 55$ kpc at 600 MHz \citep{vanweeren10}.
As shown in Figures 6 and 7, in all the models except {\bf SA4}, 
the calculated FWHM of $I_{\nu}(R)$ at 600 MHz (red dotted lines) at the acceleration age of $60-110$ Myr 
would be compatible with the observed value.
In {\bf SA4} model, the intensity profile at 600 MHz seems a bit too broad to fit
the observed profile.  

\subsection{Volume-Integrated and Postshock Spectra}

Figures 8 and 9 show the volume-integrated synchrotron spectrum, $J_{\nu}$, and its slope, $A_{\nu}$,
for the same models at the same ages as in Figures 6 and 7.
Here $J_{\nu}$ is integrated over the curved, ribbon-like downstream volume 
with $\psi=10^{\circ}$, as described in Section 3.2.
In the figures, $J_{\nu}$ is plotted in arbitrary units.
The filled circles represent the data points for the integrated flux of the Sausage relic,
which were taken from Table 1 of \citet{stroe14b} and re-scaled roughly to fit by eye the spectrum
of {\bf SA1} model at $60\Myr$ (the red dotted line in the top-left panel of Figure 8).
{\bf 
Observational errors given by \citet{stroe14b} are about 10 \%,
so the error bars are in fact smaller than the size of the filled circles in the figures,
except the one at 16~GHz with 25 \%.}

At first it looks that in most of the models, the calculated $J_{\nu}$ reasonably fits the 
observed data including the point at 16 GHz.
But careful inspections indicate that our models fail to reproduce the rather abrupt increase 
in the spectral curvature at $\sim 1.5$ GHz.
The calculated $J_{\nu}$ is either too steep at low frequencies $\la 1.5$ GHz, or
too flat at high frequencies $\ga 1.5$ GHz.
For instance, $J_{\nu}$ of the fiducial model, {\bf SA1}, is too steep at $\nu \la 1.5$ GHz.
On the other hand, {\bf SA1p} model with a flatter fossil population and {\bf SB1} model 
with a flatter injection spectrum (due to higher $M_s$) seem to have the downstream electron
spectra too flat to explain the observed data for $\nu \ga 1.5$ GHz.

As mentioned before, the transition of the integrated spectral index from 
$\alpha_{\rm inj}$ to $\alpha_{\rm inj} + 0.5$
occurs gradually over a broad range of frequency, $(0.1-10) \nu_{\rm br}$.
If we estimate $\nu_{\rm br}$ as the frequency at which the gradient, $d A_{\nu} /d \nu$, has the 
largest value, it is $\sim 4$, 1, 0.3 GHz at $t_{\rm age} = 30,\ 60,\ {\rm and}\ 110\Myr$,
respectively, for all the models except {\bf SC1pex1} model (for which $A_{\nu}$ is
shown at different epochs).
Since $\nu_{\rm br}$ is determined mainly by the magnetic field strength and the acceleration age, 
it does not sensitively depend on other details of the models.
{ 
In {\bf SC1pex1} model, the power-law portion ($\propto p^{4.2}$) gives 
a flatter spectrum with $A_{\nu}\approx 0.7$ at low frequencies, while the newly injected population 
at the shock with $M_s\approx 2.1$ results in a steeper spectrum with $A_{\nu}\approx 1.6$ at high frequencies.
Note that the spectral index estimated with the observed flux in \citet{stroe14b} 
between 2.3 and 16 GHz is $A_{\nu}\approx 1.62$, implying $M_s\approx 2.1$.
}

In Figure 10, $J_{\nu}$'s from all the models considered here are compared 
at $t_{\rm age} = 60$ Myr ({\bf SA} and {\bf SB} shocks) or $80$ Myr ({\bf SC} shocks).
The observed data points of the Sausage relic are also shown.
{
In most of the models ({\bf SA} models), the shocks have $M_s\approx 3.0-3.1$ at 60 Myr,
so the predicted $J_{\nu}$ between 2.3 and 16 GHz is a bit flatter with $A_{\nu} \approx 1.25$ 
than the observed spectrum with $A_{\nu}\approx 1.62$ at the same frequency range.}
As mentioned above, {\bf SA1p} and {\bf SB1} models produce $J_{\nu}$, which is significantly flatter 
at high frequencies than the observed spectrum.
{ 
The toy model, {\bf SC1pex1}, seems to produce the best fit to the observed spectrum,
as noted before.
}

In short, the volume-integrated synchrotron spectra calculated here
steepen gradually over $(0.1-10) \nu_{\rm br}$ with the break frequency $ \nu_{\rm br}\sim 1$ GHz,
if $t_{\rm age} \sim 60 - 80$ Myr. 
However, all the models considered here seem to have some difficulties fitting the sharp curvature 
around $\sim 1.5$ GHz in the observed integrated spectrum. 
This implies that the shock dynamics and/or the downstream magnetic field profile could be different
from what we consider here.
Perhaps some additional physics that can introduce a feature in the electron energy 
spectrum, stronger than the `$q+1$' steepening due to radiative cooling, might be necessary.

In Figure 11, we present for the three models, {\bf SA1}, {\bf SA1b}, and {\bf SC1pex1}, 
the mean intensity spectrum in the downstream regions of 
$[R_i, R_i+5{\rm kpc}]$, $\langle I_\nu \rangle = \int_{R_i}^{R_i+5} I_\nu(R) dR$,
where $R_i= 5\kpc \cdot (2i-1)$ and $i$ runs from 1 to 6.
This is designed to be compared with the `spatially resolved' spectrum behind the shock in radio observations
\citep[e.g.,][]{stroe13}.
The figure shows how the downstream spectrum cuts off at progressively lower frequencies
due to radiative cooling as the observed position moves further away from the shock.

\section{SUMMARY}

{ 
We propose a model that may explain some characteristics of giant radio relics: 
the relative rareness, 
uniform surface brightness along the length of thin arc-like radio structure, 
and spectral curvature in the integrated radio spectrum over $\sim (0.1-10)$ GHz.
In the model, a spherical shock encounters an elongated cloud of the ICM thermal gas that is permeated 
by enhanced magnetic fields and an additional population of fossil relativistic electrons.
As a result of the shock passage, the fossil electrons are re-accelerated to radio-emitting 
energies ($\gamma_e \la 10^4$), resulting in a birth of a giant radio relic. 
}\

In order to explore this scenario,
we have performed time-dependent, DSA simulations of spherical
shocks with the parameters relevant for the Sausage radio relic in cluster CIZA
J2242.8+5301.
In the fiducial model, the shock decelerates from $u_{s}\approx 3.3\times 10^3 \kms$ ($M_s\approx 3.5$)
to $u_{s}\approx 2.9\times 10^3 \kms$ ($M_s\approx 3.1$) at the acceleration age of 60 Myr.
The seed, fossil electrons with $\gamma_{\rm e,inj}\sim 30$ are assumed to be injected into
the CR population, which is subsequently re-accelerated to higher energies.
Such shocks are expected to produce the electron energy spectrum, $f_e(p)\propto p^{-4.5}$,
resulting in the synchrotron radiation spectrum with the injection index, $\alpha_{\rm inj}\approx 0.75$,
and the integrated index, $A_{\nu} \approx 1.25$, at high frequencies ($\ga 1$ GHz).
We consider various models with a range of shock parameters, different upstream gas density profiles,
different downstream magnetic field profiles, and three types of fossil electron populations,
as summarized in Table 1.
Adopting a ribbon-like curved shock surface and the associated downstream volume,
which are constrained by the extension angle (or viewing depth) of $\psi=10^{\circ}$ as detailed in
Section 2.5 \citep[e.g][]{vanweeren10,kang12},
the radio surface brightness profile, $I_{\nu}(R)$, and the volume-integrated spectrum, 
$J_{\nu}$, are calculated.

The main results are summarized as follows.

{ 
1) Two observables, the break frequency in the integrated synchrotron spectrum, $\nu_{\rm br}$,
and the width of the synchrotron emission region behind the shock, $\Delta l_{\nu_{\rm obs}}$, 
can have identical values for two values of postshock magnetic field strength
(see Equations [\ref{fbr}] and [\ref{lwidth}]).

2) The observed width of the surface brightness projected onto the sky plane,
$\Delta l_{\rm SB}$, strongly depends on the assumed value of $\psi$ (see Figure 4).
So $\Delta l_{\rm SB}$ may not be used to estimate
the postshock magnetic field strength, unless the projection effects can be modeled properly.

3) The integrated synchrotron spectrum is expected to have a spectral curvature that runs over
a broad range of frequency, typically for $(0.1-10)\nu_{\rm br}$.
For a shock of $M_s \approx 3$ with the postshock magnetic field strength, $B_{\rm low} \sim 0.62 \muG$ or  
$B_{\rm high} \sim6.2\muG$, the integrated spectral index 
increases gradually from $\alpha_{\rm inj}\approx 0.75$ to $\alpha_{\rm inj}+0.5\approx 1.25$ 
over $0.1-10$ GHz, if the duration of the shock acceleration is $\sim 60$ Myr.

4) Assuming that the upstream sound speed is $c_{s,1}\approx 920\kms$ ($kT_1\approx 3.35$~keV)
as inferred from X-ray observation,
a shock of $M_s \approx 3$ and $u_s \approx 3\times 10^3 \kms$ (e.g., {\bf SA1} model) can reasonably 
explain the observed width, $\Delta l_{\rm SB}\sim 55$~kpc \citep{vanweeren10}, and 
the curved integrated spectrum of the Sausage relic \citep{stroe14b}.
{\bf SB1} model with a shock of $M_s \approx 4.5$, however, produces the integrated spectrum that 
seems too flat to explain the observed spectrum above $\sim 1$~GHz.

5) We also consider two toy models with
power-law electron populations with exponential cutoffs at $\gamma_e\sim 10^4$,
$f_{\rm e,up}(p)\propto p^{-4.2}\exp[-(\gamma_e/\gamma_{e,cut})^2]$ ({\bf SC1pex1} and {\bf SC1pex2} models).
They may represent the electron populations that were produced
earlier and then have cooled down to $\gamma_e\sim 10^4$.
{\bf SC1pex1} model with a weaker shock ($M_s\approx 2.1$) reproduces better the characteristics 
of the observed integrated spectrum.
But the steepening of the integrated spectrum due to radiative cooling alone
may not explain the strong spectral curvature above 1.5 GHz toward 16 GHz. 
} 

6) This strong curvature at $\sim 1.5$ GHz may imply that the downstream electron energy 
spectrum is influenced by some additional physical 
processes other than radiative losses, because the integrated spectrum of radiatively cooled 
electrons steepens with the frequency only gradually.
This conclusion is likely to remain unchanged even in the case where the observed spectrum consists of the synchrotron emission from multiple shocks with different Mach numbers,
as long as the postshock electrons experience only simple radiative cooling.

{ 
Other models that may explain the curved spectrum will be further explored and presented elsewhere.
}
\acknowledgements

{ 
The authors thank the anonymous referee for his/her thorough review and constructive suggestions that lead to
a significant improvement of the paper.}
HK was supported by Basic Science Research Program through the National Research Foundation of Korea (NRF) funded by the Ministry of Education (2014R1A1A2057940).
DR was supported by the National Research Foundation of Korea through grant NRF-2014M1A7A1A03029872 and NRF-2012K1A3A7A03049606.

\clearpage

\begin{deluxetable} {lccccccc}
\tablecaption{Parameters for DSA Simulations}
\tablehead{
\colhead {Model} & \colhead{$\rho_{\rm up}(r)$}  & \colhead{$B_1\ ^a$}& \colhead{$B_2(t)\ ^{a,b}$} & \colhead{$kT_1$} 
& \colhead{$u_{s}(t)\ ^b$} & \colhead{$M_{s}(t)\ ^b$} & \colhead{$f_{\rm e,up}(p)$}\\
\colhead {Name} & \colhead{ }  & \colhead{ $(\mu{\rm G})$} & \colhead{ $(\mu{\rm G})$} & \colhead{keV} 
& \colhead{$( 10^3\kms)$} & \colhead{ } & \colhead{}
}
\startdata
{\bf SA1}    & $\propto r^{-2}$ & 2.5  & $6.7-6.3$   &  $3.35$ & $3.3-2.9$ & $3.5-3.1$ & $-$  \\
{\bf SA1b}   & $\propto r^{-2}$ & 0.25 & $0.67-0.63$ & $3.35$ & $3.3-2.9$ & $3.5-3.1$ & $-$ \\
{\bf SA1p}   & $\propto r^{-2}$ & 2.5  & $6.7-6.3$ &    $3.35$ & $3.3-2.9$ & $3.5-3.1$ & $\propto p^{-4.2}$ \\
{\bf SA2}$^c$ & $\propto r^{-2}$ & 2.0 & $7\muG$ &  $3.35$ & $3.3-2.9$ & $3.5-3.1$ & $-$ \\
{\bf SA3}    & $\propto r^{-4}$ & 2.5  & $6.7-6.3$    &  $3.35$ & $3.0-2.8$ & $3.2-3.0$ & $-$  \\
{\bf SA4}$^c$  & $\rho_0$  & 2.0 &      $7\muG$       & $3.35$ & $3.3-2.8$ & $3.5-2.9$ & $-$ \\ 
{\bf SB1}   & $\propto r^{-2}$ & 2.5   & $7.5-7.3$     & $1.49$ & $3.3-2.9$ & $5.3-4.6$ & $-$ \\
{\bf SC1pex1}$^d$ & $\propto r^{-2}$ & 2.5 &  $5.5-5.1$ & $3.35$ & $2.2-1.9$ & $2.4-2.1$ & $\propto p^{-4.2}\cdot \varepsilon_{c1}$ \\
{\bf SC1pex2}$^e$ & $\propto r^{-2}$ & 2.5 &  $5.5-5.1$ & $3.35$ & $2.2-1.9$ & $2.4-2.1$ & $\propto p^{-4.2}\cdot \varepsilon_{c2}$ \\
\enddata
\tablenotetext{a}{$B_1$ and $B_2$ are magnetic field strengths immediately upstream and 
downstream of the shock, respectively. See Equation (\ref{b2}).} 
\tablenotetext{b}{Since the shock is decelerating, the ranges of values are given for the acceleration age 
$t_{\rm age} = 0- 60 \Myr$. }
\tablenotetext{c}{The downstream ($r<r_s$) magnetic field strength, $B_{\rm dn} = B_2$ in {\bf SA2} and {\bf SA4} models,
while $B_{\rm dn}(r,t)= B_2(t) \cdot [P_g(r,t)/P_{g,2}(t)]^{1/2}$ in the rest of the models.}
\tablenotetext{d}{$\varepsilon_{c1}=\exp[-(\gamma_e/10^4)^2]$. }
\tablenotetext{e}{$\varepsilon_{c2}=\exp[-(\gamma_e/2\times10^4)^2]$. }
\end{deluxetable}

\clearpage

\begin{figure}
\vspace{1cm}
\hskip 1.5cm
\includegraphics[scale=0.4]{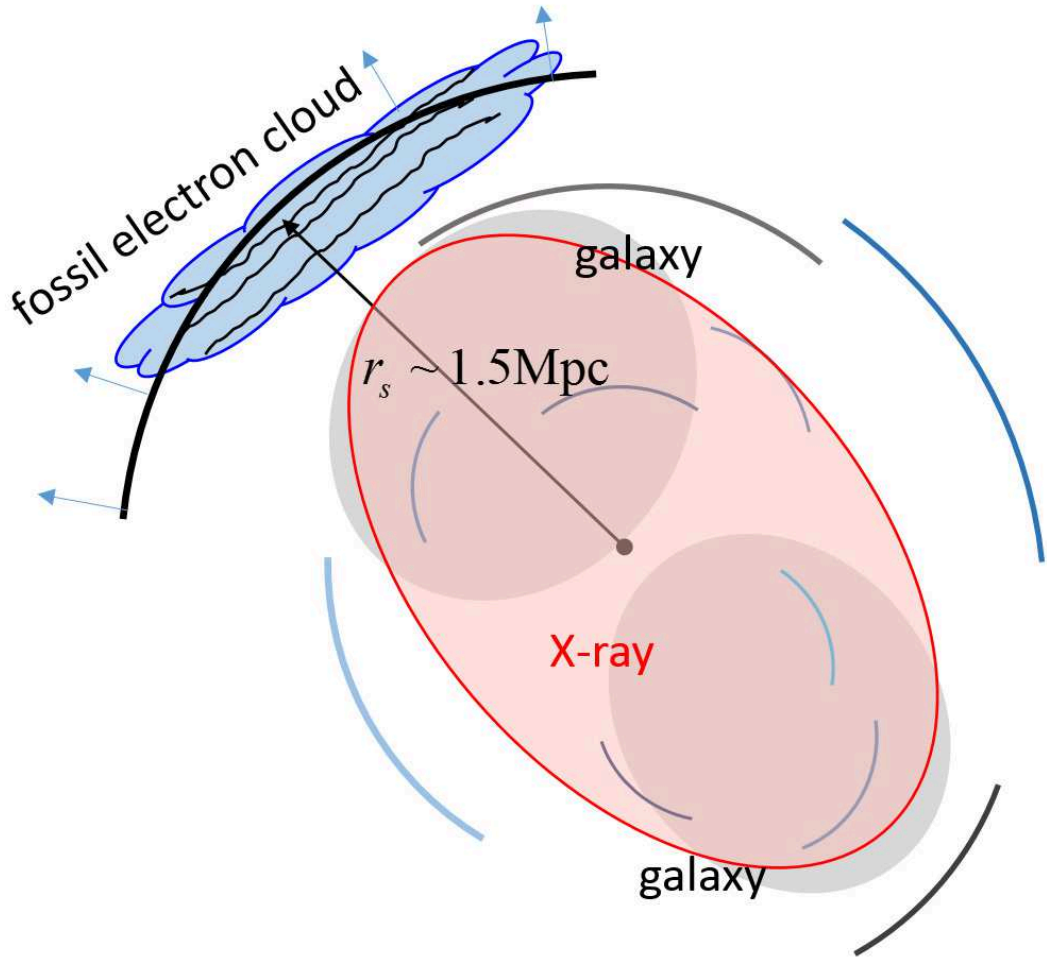}
\caption{Schematic diagram illustrating a scenario for the Sausage relic in a merging cluster,
CIZA J2242.8+5301:
a shock impinges on a fossil electron cloud with regular magnetic field in the cluster outskirt.
{ Two gray disks represent the galaxy distribution, while the red ellipse shows the distribution of
X-ray emitting gas. Arcs are meant to depict abundant shocks with different Mach numbers that
are expected to form during a merger event.} }
\label{Fig1}
\end{figure}

\clearpage

\begin{figure}
\vspace{-1cm}
\hskip -0.8cm
\includegraphics[scale=0.87]{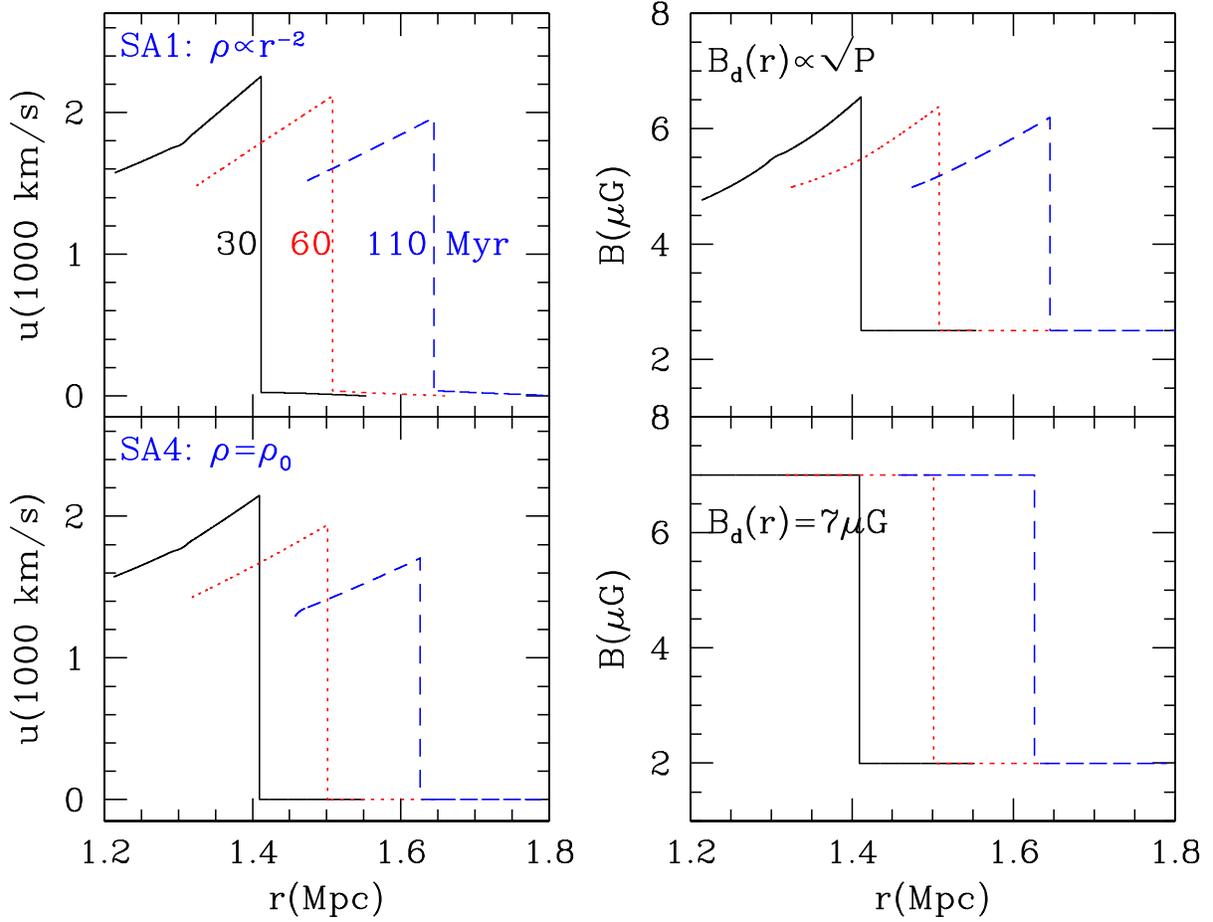}
\vspace{-4.5cm}
\caption{Time evolution of the flow speed, $u(r,t)$, and the magnetic field, $B(r,t)$,
of the spherical shock in {\bf SA1} (top two panels) and {\bf SA4} (bottom two panels) models
at the acceleration age, $t_{\rm age}=30$, 60, 110 Myr (black solid, red dotted, and blue dashed lines, respectively).}
\label{Fig2}
\end{figure}

\clearpage

\begin{figure}
\vspace{1cm}
\hskip 1.5cm
\includegraphics[scale=0.78]{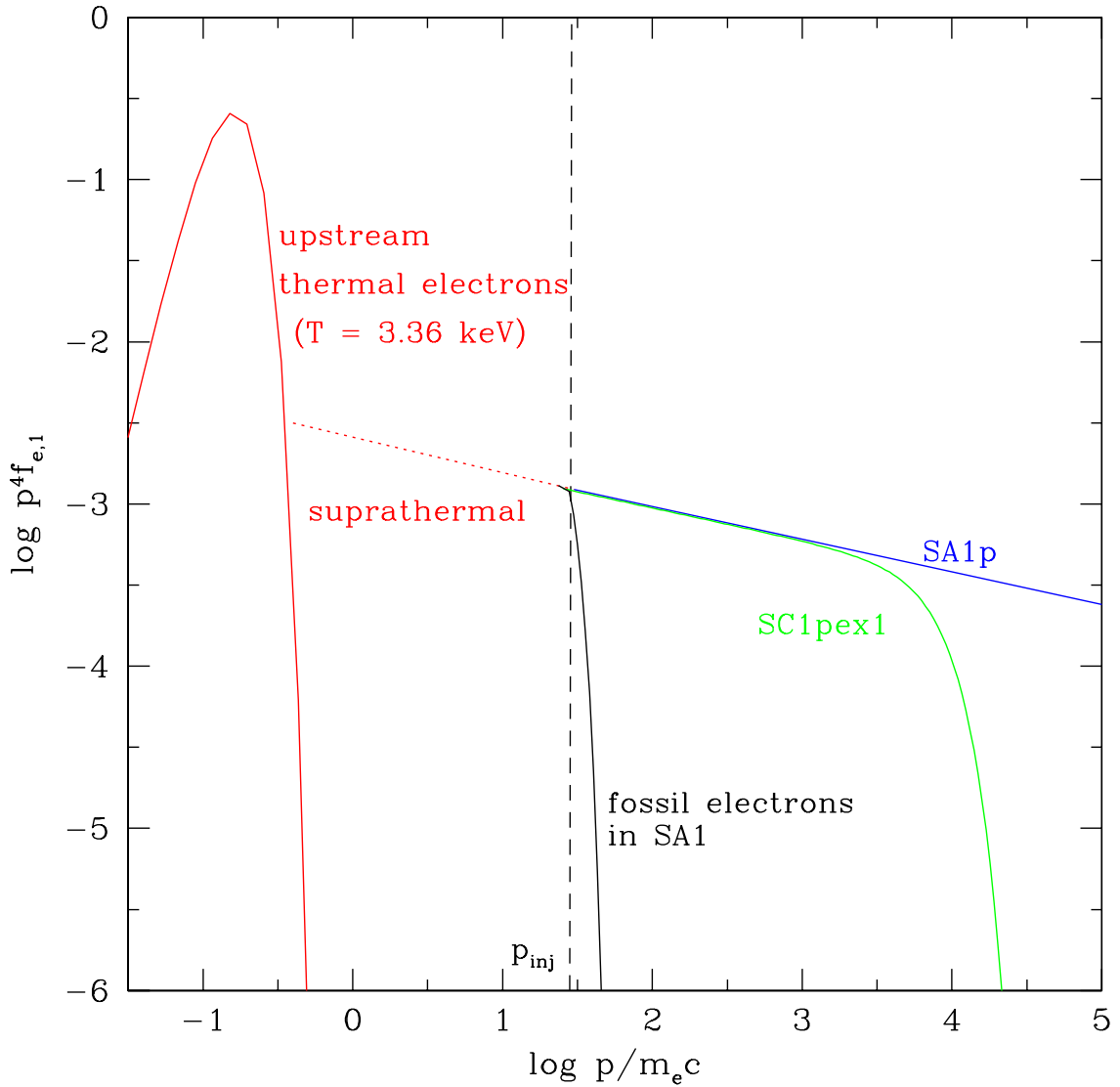}
\caption{Electron distributions in the cloud medium: the Maxwellian distribution of thermal
electrons with $kT=3.35$ keV (red solid line),
fossil electrons with $\gamma_e\approx 30$ in {\bf SA1} model (black solid),
a power-law of $p^{-4.2}$ in {\bf SA1p} model (blue solid), 
and a power-law with an exponential cutoff in {\bf SC1pex1} model (green solid).
See Table 1 for the different model parameters.
The vertical dashed line demarcates the injection momentum,
$p_{\rm inj}$. 
An unspecified suprathermal distribution is shown in the red dotted line, but
the electron distribution below $p_{\rm inj}$ is not relevant for DSA.
}
\label{Fig3}
\end{figure}

\clearpage

\begin{figure}
\includegraphics[scale=0.78]{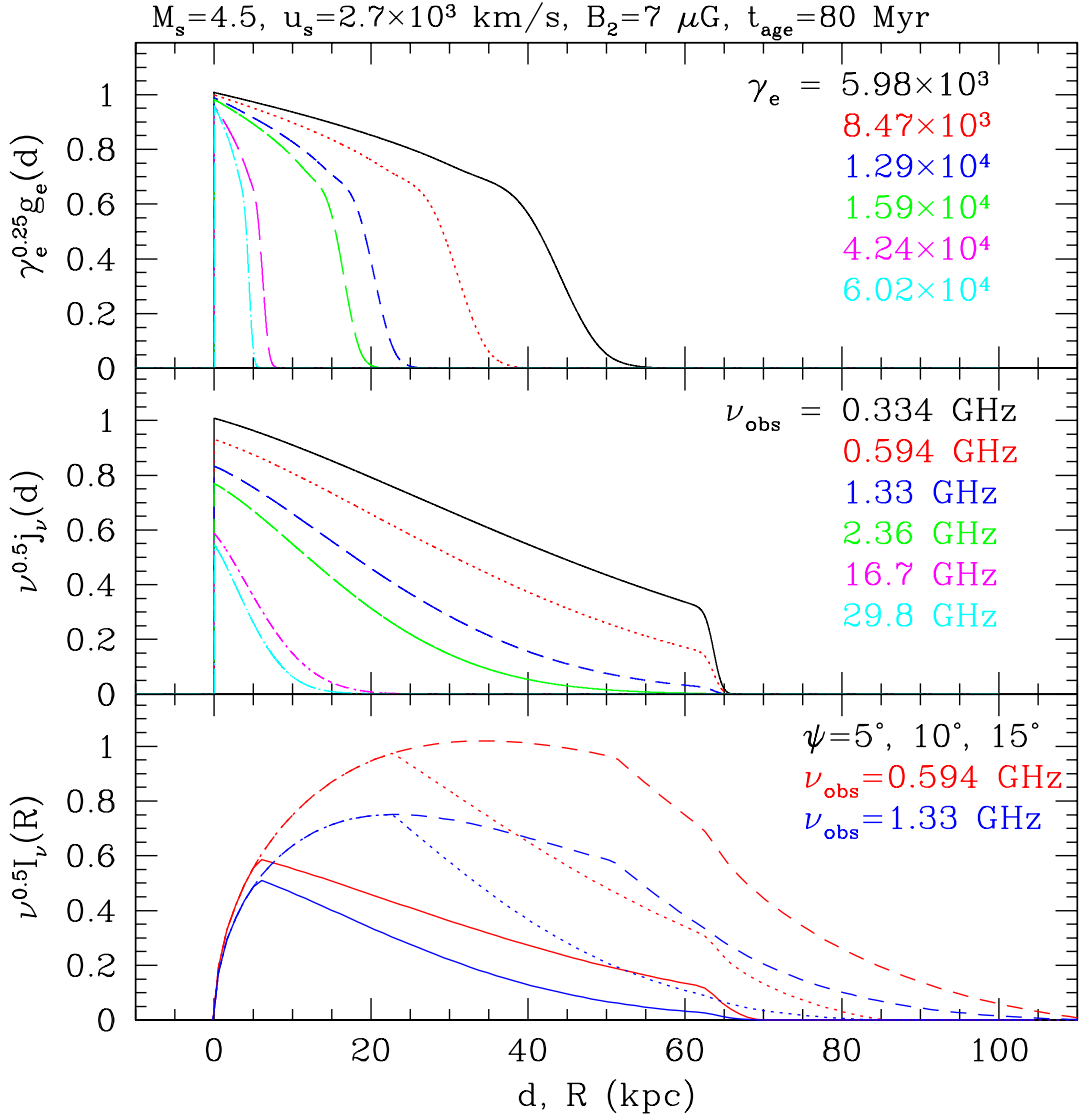}
\vspace{-1.0cm}
\caption{Spatial profiles of $g_e(d)$, $j_{\nu}(d)$, and $I_{\nu}(R)$ for a steady {\it planar} shock
with  $u_s=2.7\times 10^3 \kms$ and $M_s=4.5$ at the acceleration age of 80 Myr.
Here, $d$ is the downstream distance away from the shock, while $R$ is the distant behind the shock
projected in the sky plane.
Top: $\gamma_e^{0.25}g_e(d)$ is plotted for the electron Lorentz factor, $\gamma_e= 5.98\times 10^3$ 
(black solid line), $8.47\times 10^3$ (red dotted), $1.29\times 10^4$ (blue dashed), $1.59\times 10^4$ 
(green long dashed), $4.24\times 10^4$ (magenta dot-dashed), and $6.02\times 10^4$ (cyan dot-long dashed).
Middle: $\nu^{0.5}j_{\nu}(d)$ is plotted for the observation frequency, $\nu_{\rm obs}= 0.334$ GHZ 
(black solid line), 0.594 GHz (red dotted), 1.33 GHz (blue dashed), 2.36 GHz (green long dashed), 
16.7 GHz (magenta dot-dashed), and 29.8 GHz (cyan dot-long dashed).
Bottom: $\nu^{0.5}I_{\nu}(R)$ is plotted for the extension angle 
$\psi= 5^{\circ},\ 10^{\circ}$, and $15^{\circ}$ (solid, dotted, and dashed lines, respectively).
The red (blue) curves are for $\nu_{\rm obs}=0. 594\ {\rm GHz}$ (1.33 GHz).
The redshift of the host cluster is assumed to be $z=0.192$ here.}
\label{Fig4}
\end{figure}

\clearpage

\begin{figure}
\vspace{-0.5cm}
\hskip -0.2cm
\includegraphics[scale=0.8]{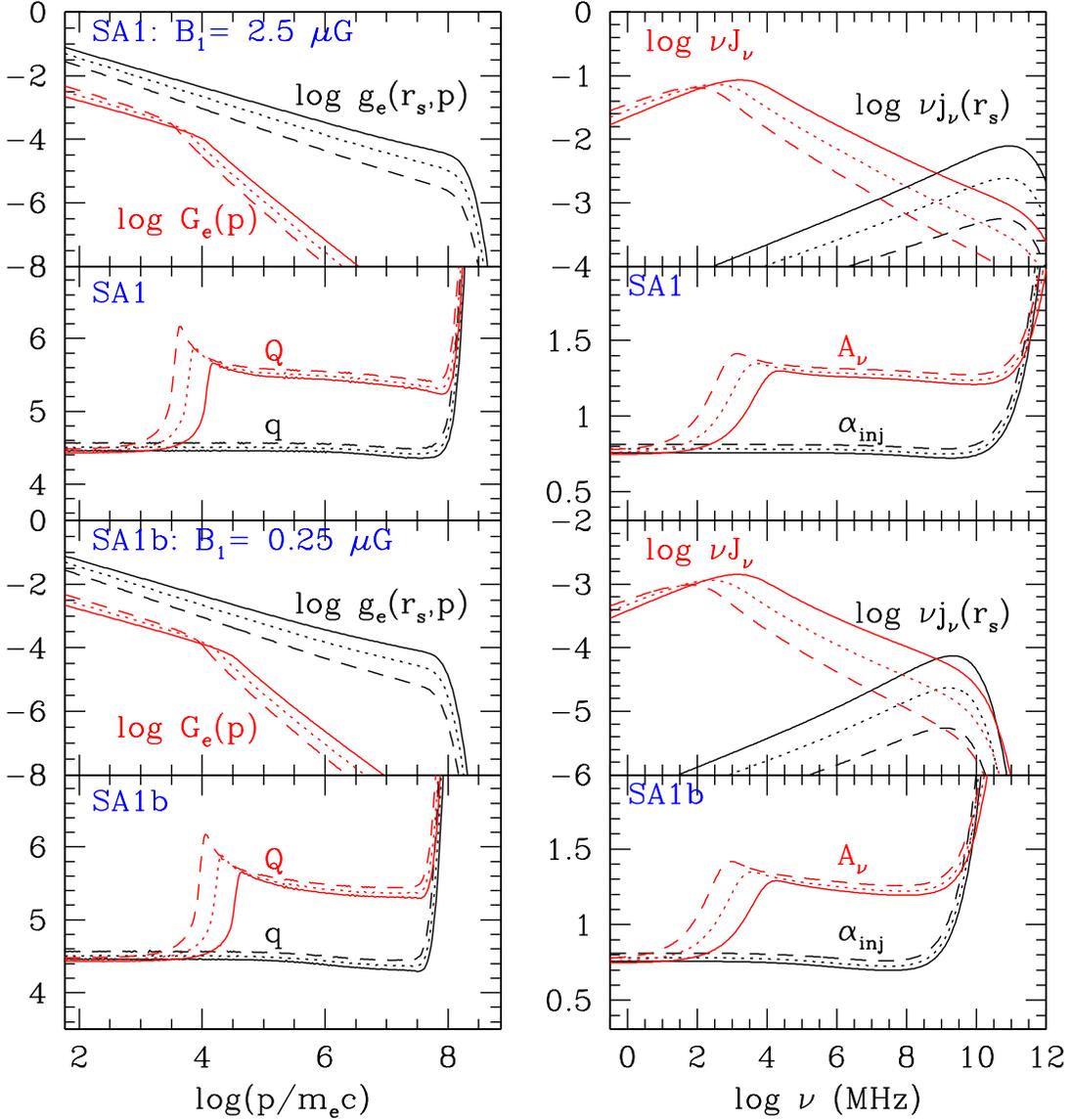}
\vspace{-0.5cm}
\caption{
Comparison of {\bf SA1} model with $B_1=2.5 \muG$ (upper four panels) and 
{\bf SA1b} model with $B_1=0.25 \muG$ (lower four panels)
at $t_{\rm age}=30$, 60, 110 Myr (solid, dotted, and dashed lines, respectively).
Left: Electron distribution function at the shock position, $g_e(r_s,p)=p^4 f_e(r_s,p)$ (black lines),
volume-integrated electron distribution function, $G_e(p)=p^4 F_e(p)= 
p^4 \int f_e(r,p) dV$ (red lines), and
slopes of electron distribution functions, $q=-d \ln f_e(r_s)/d \ln p$ (black lines) and
$Q=-d \ln F_e/d \ln p$ (red lines).
Right: Synchrotron spectrum at the shock position, $\nu j_{\nu}(r_s)$ (black lines),
volume-integrated synchrotron spectrum, $\nu J_{\nu}= \nu \int j_{\nu}(r) dV $ (red lines), and
synchrotron spectral indices, $\alpha_{\rm inj}= -d \ln j_{\nu}(r_s)/d \ln \nu $ (black lines)
and $A_{\nu}= -d \ln J_{\nu}/d \ln \nu $ (red lines).}
\label{Fig5}
\end{figure}

\clearpage

\begin{figure}
\vspace{-0.5cm}
\hskip -0.3cm
\includegraphics[scale=0.85]{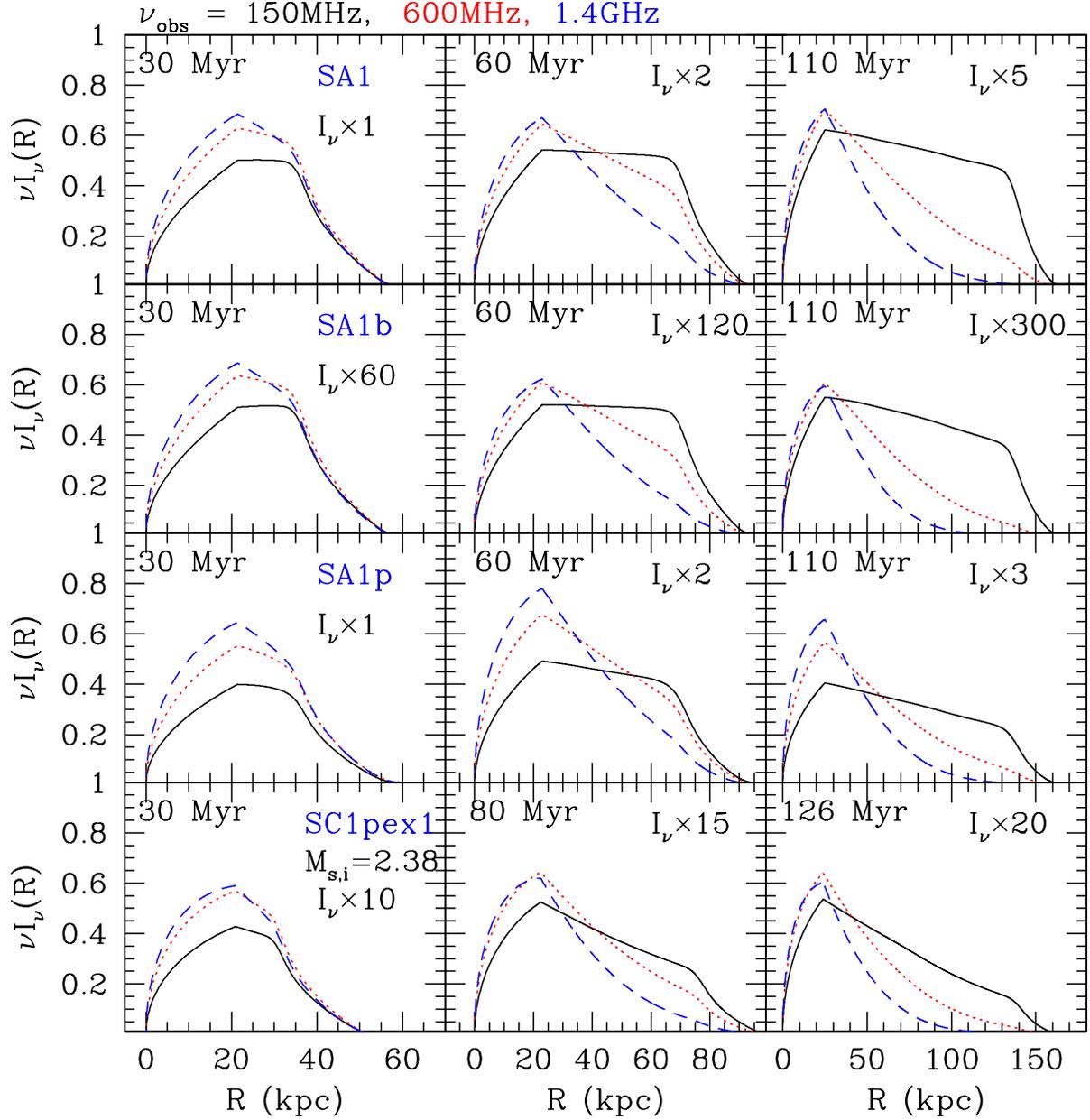}
\vspace{-0.8cm}
\caption{Surface brightness profiles at $\nu_{\rm obs}=150$, 600, and 1400 MHz
(black solid, red dotted, blue dashed lines) in {\bf SA1}, {\bf SA1b}, {\bf SA1p}, 
and {\bf SC1pex} models (from top to bottom panels).
See Table 1 for the model parameters.
The results are shown at the acceleration age, $t_{\rm age}$, specified in each panel.
For the extension angle, $\psi = 10^{\circ}$ is adopted.
The quantity $\nu I_{\nu} \times X$ is plotted with a scale factor $X$ 
to present it in the same ordinate scale
for all the models.}
\label{Fig6}
\end{figure}

\clearpage

\begin{figure}
\vspace{-0.5cm}
\hskip -0.3cm
\includegraphics[scale=0.85]{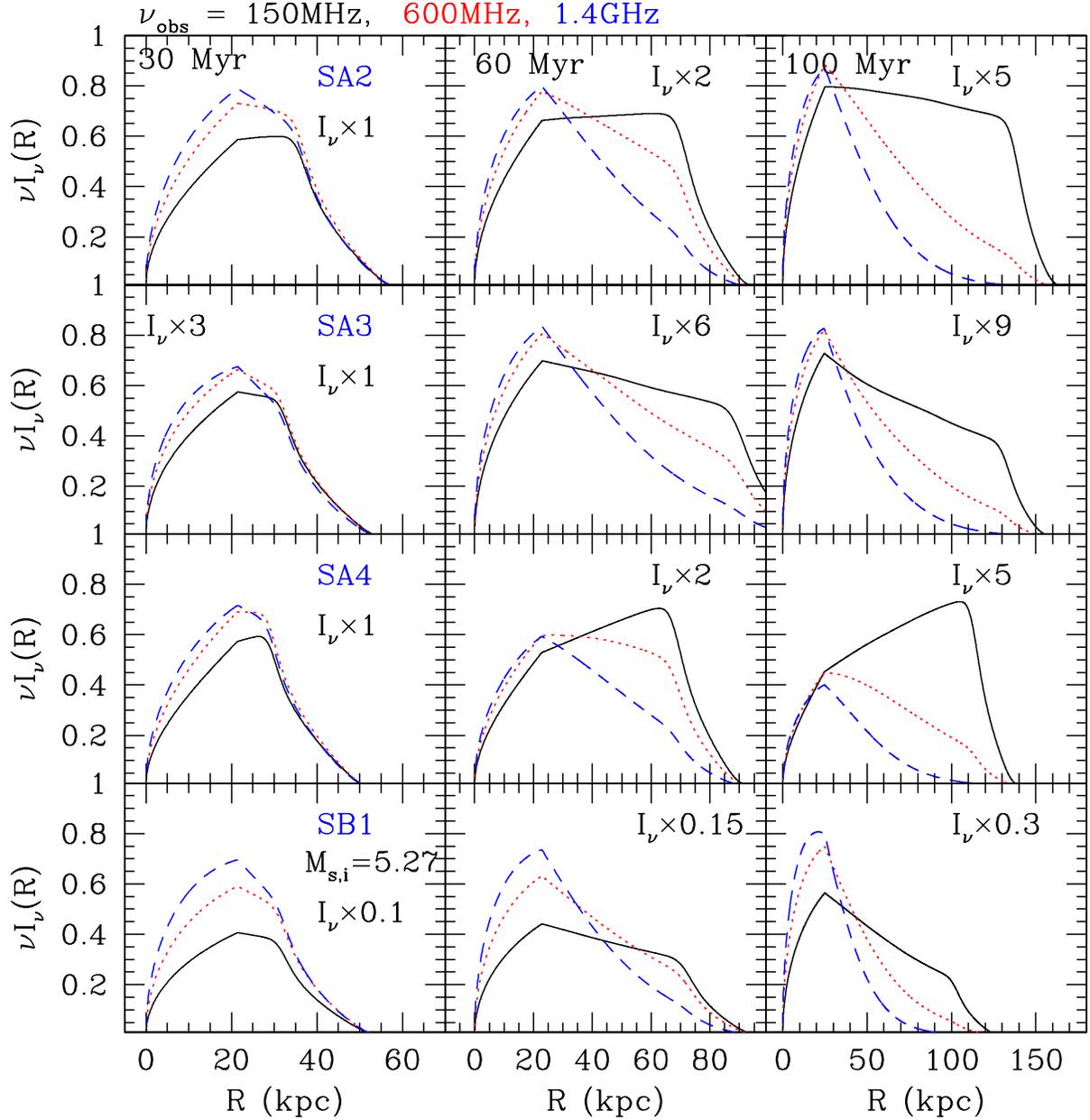}
\vspace{-0.8cm}
\caption{Surface brightness profiles at $\nu=150$, 600, and 1400 MHz
(black solid, red dotted, blue dashed lines) in {\bf SA2}, {\bf SA3}, {\bf SA4}, 
and {\bf SB1} models (from top to bottom panels).
See Table 1 for the model parameters.
The results are shown at $t_{\rm age} =$ 30, 60, and 110 Myr (from left to right panels).
For the extension angle, $\psi = 10^{\circ}$ is adopted.
The quantity $\nu I_{\nu} \times X$ is plotted with a scale factor $X$ to present
it in the same ordinate scale for all the models.}
\label{Fig7}
\end{figure}

\clearpage

\begin{figure}
\vspace{-0.5cm}
\hskip -0.3cm
\includegraphics[scale=0.85]{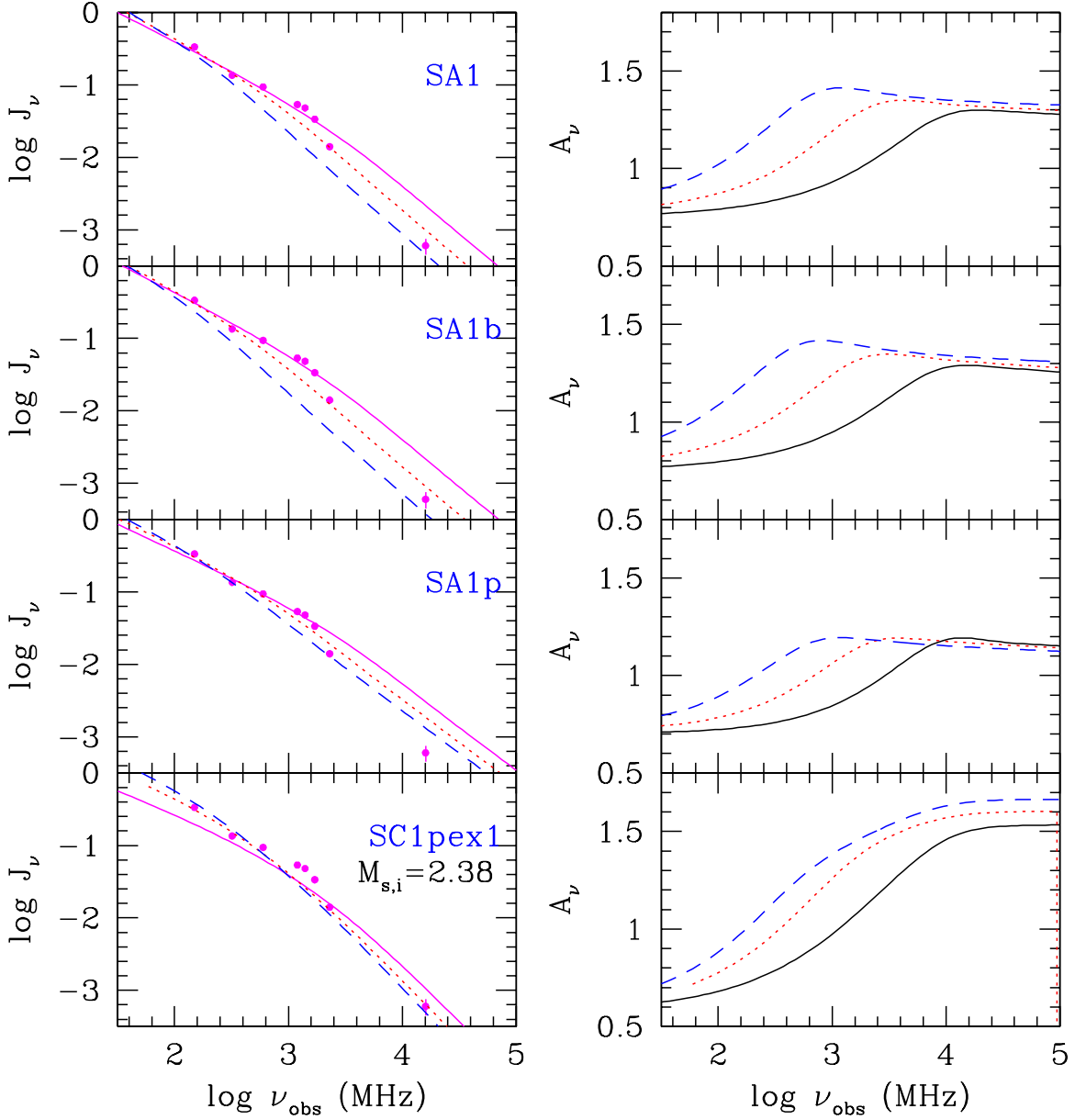}
\vspace{-1.1cm}
\caption{Volume-integrated synchrotron spectrum, $J_{\nu}$, and its spectral index, $A_{\nu}$,
at $t_{\rm age} = 30$, 60, and 110 Myr (black solid, red dotted, and blue dashed lines, respectively)
for {\bf SA1}, {\bf SA1b}, and {\bf SA1p} models (top three panels), and
at $t_{\rm age} = 30$, 80, and 126 Myr (black solid, red dotted, and blue dashed lines, respectively)
for {\bf SC1pex1} model (bottom panel).
Filled circles show the data from \citet{stroe14b}, scaled to fit by eye the red dotted line
($t_{\rm age} = 60 \Myr$) for {\bf SA1} model.
Observational errors are small, about $10 \%$, except for the data at 16~GHz with $25\%$
(shown in a vertical bar).}
\label{Fig8}
\end{figure}

\clearpage

\begin{figure}
\vspace{-0.5cm}
\hskip -0.3cm
\includegraphics[scale=0.85]{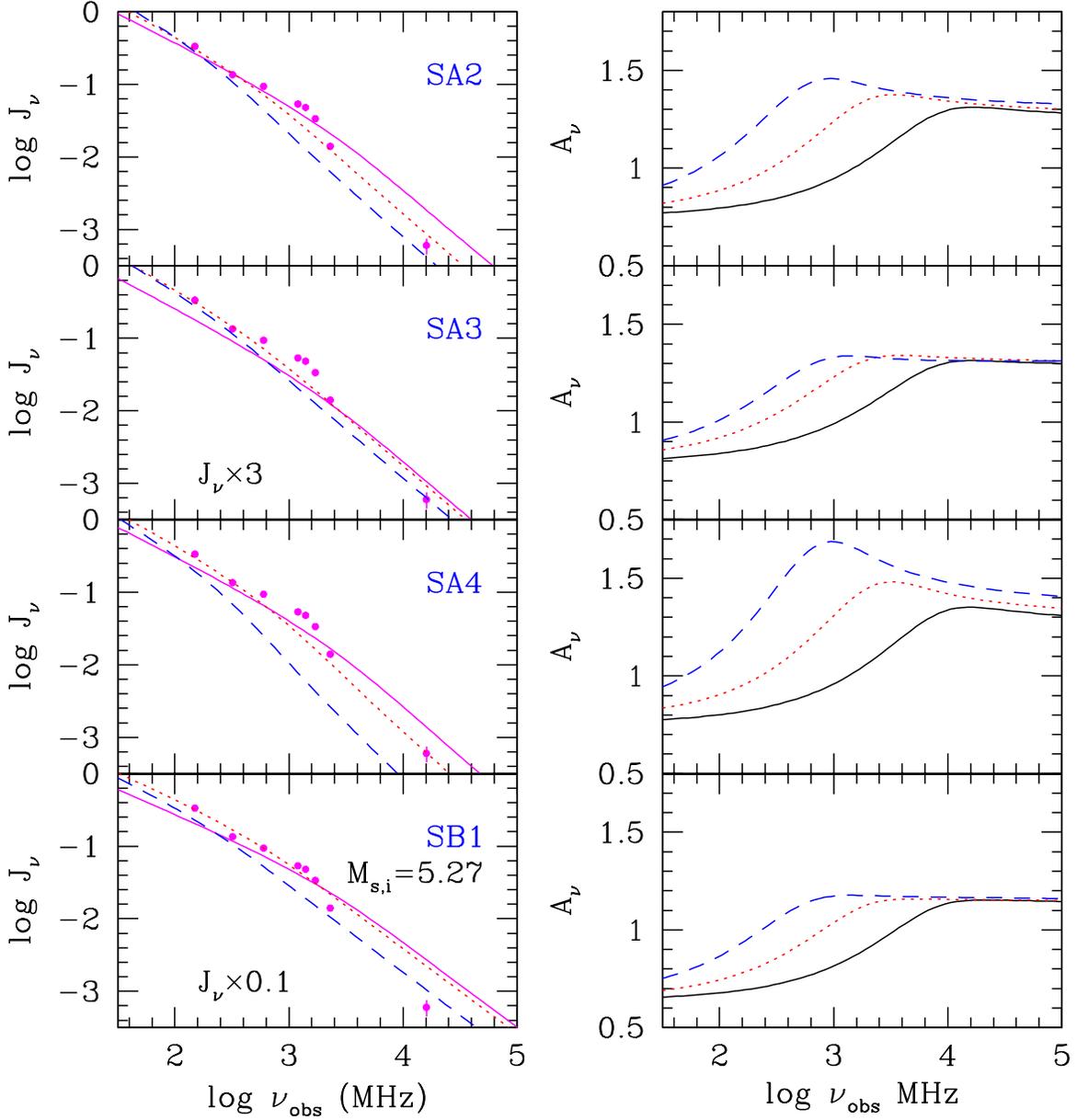}
\vspace{-1.1cm}
\caption{Volume-integrated synchrotron spectrum, $J_{\nu}$, and its spectral index, $A_{\nu}$,
at $t_{\rm age} = 30$, 60, and 110 Myr (black solid, red dotted, and blue dashed lines, respectively)
for {\bf SA2}, {\bf SA3}, {\bf SA4}, and {\bf SB1} models.
Filled circles show the data from \citet{stroe14b}, scaled as in Figure 8.
}
\label{Fig9}
\end{figure}

\clearpage

\begin{figure}
\vspace{-0.5cm}
\hskip -1.2cm
\includegraphics[scale=0.9]{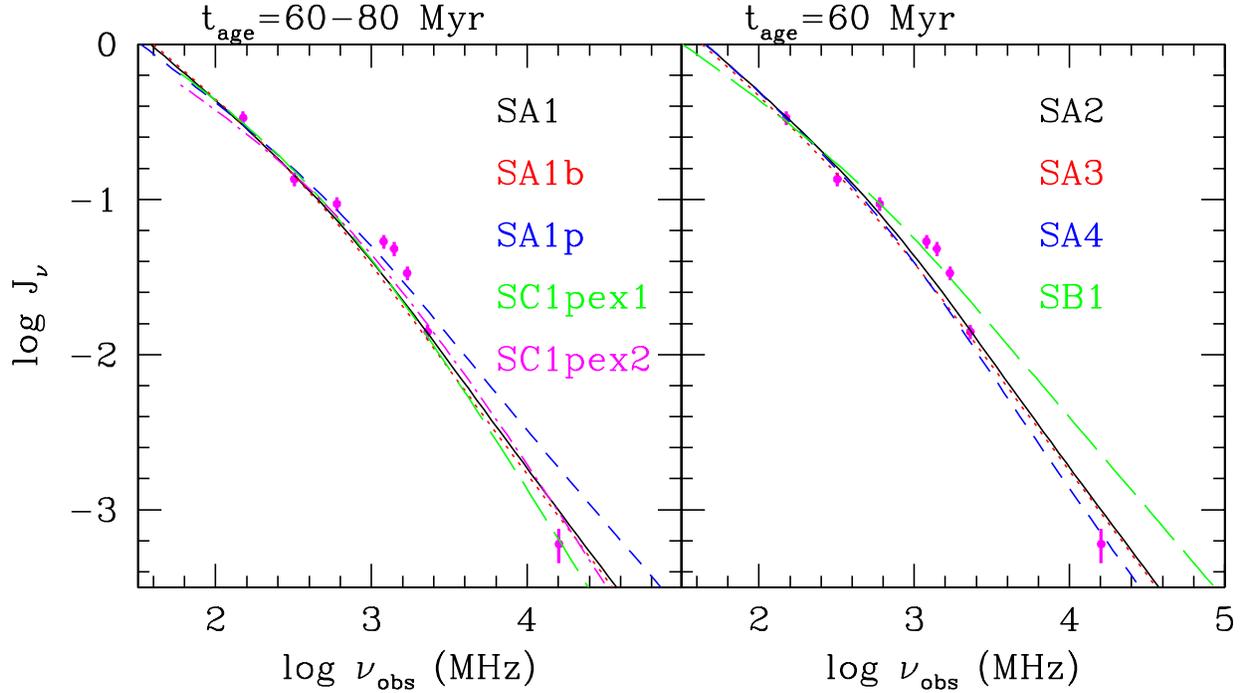}
\vspace{-8.5cm}
\caption{Volume-integrated synchrotron spectrum, $J_{\nu}$, for all the models in Table 1.
In the left panel, {\bf SA1} (black solid line), {\bf SA1b} (red dotted), {\bf SA1p} (blue dashed), 
{\bf SC1pex1} (green long dashed), and {\bf SC1pex2} (magenta dot-dashed) models are shown.
In the right panel, {\bf SA2} (black solid line), {\bf SA3} (red dotted), {\bf SA4} (blue dashed),
and {\bf SB1} (green long dashed) models are shown.
The results are shown at 60 Myr, except for {\bf SC1pex1} and {\bf SC1pex2} models
for which the results are shown at 80 Myr.
Filled circles show the data from \citet{stroe14b}, scaled as in Figure 8.
Observational errors (vertical bars) are small, about $10 \%$, except for the 
data at 16~GHz with $25\%$.}
\label{Fig10}
\end{figure}

\clearpage

\begin{figure}
\vspace{-0.5cm}
\hskip -0.3cm
\includegraphics[scale=0.85]{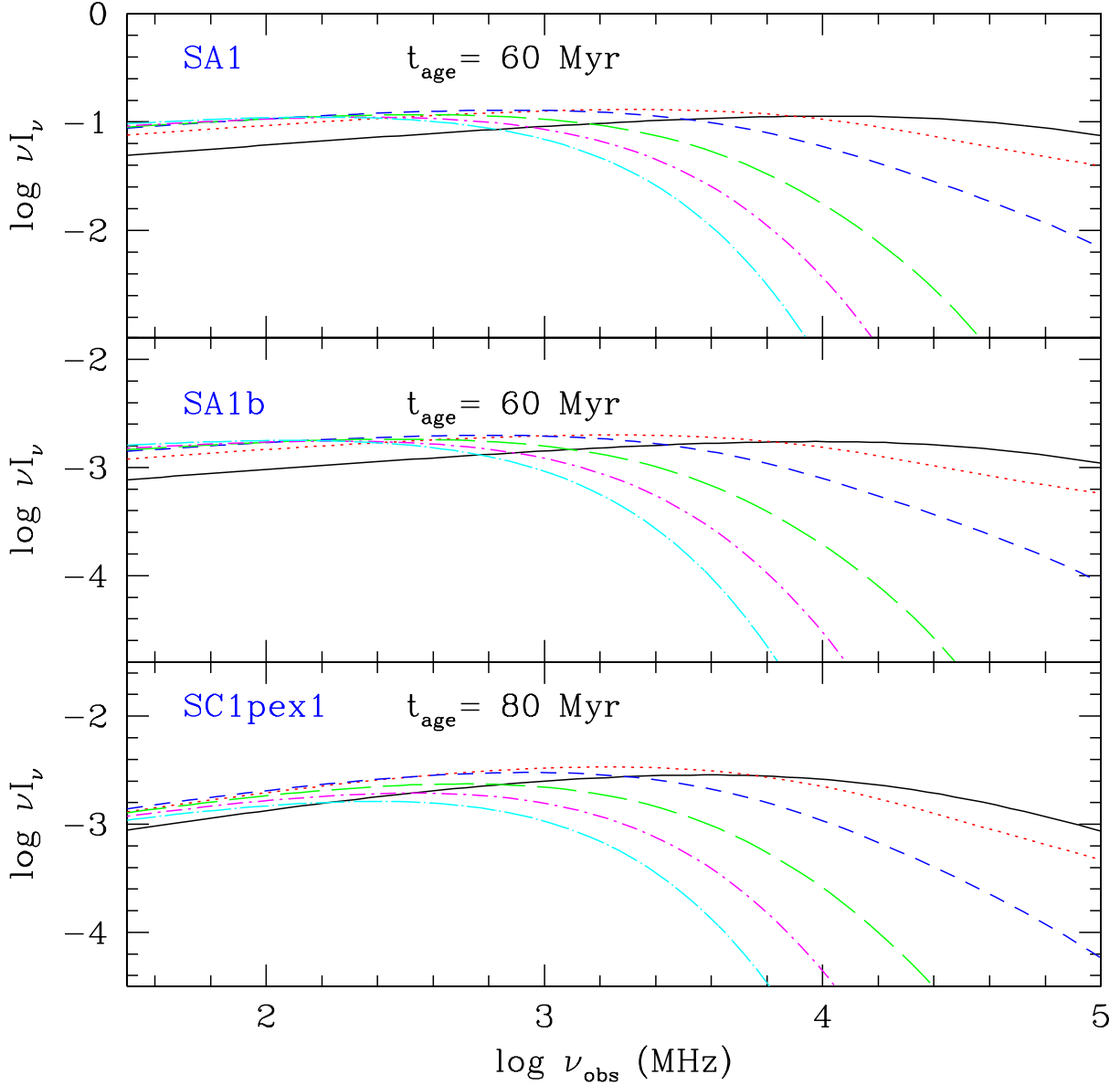}
\vspace{-1.1cm}
\caption{Mean intensity spectrum, $\langle I_{\nu} \rangle$, averaged over $[R_i,R_i+5\kpc]$\
behind the shock, where $R_i= 5\kpc \cdot (2i-1)$, $i =$ 1, 2, 3, 4, 5, 6
(black solid, red dotted, blue dashed, green long dashed, magenta dot-dashed, cyan dot-long dashed lines, respectively)
for {\bf SA1}, {\bf SA1b} and {\bf SC1pex1} models.
The results are shown at the acceleration age, $t_{\rm age}$, specified in each panel.}
\label{Fig11}
\end{figure}

\end{document}